\documentclass[a4paper,fleqn]{cas-sc}
\usepackage{amsmath}
\usepackage[numbers]{natbib}
\usepackage{caption}
\usepackage{setspace}

\def\tsc#1{\csdef{#1}{\textsc{\lowercase{#1}}\xspace}}
\tsc{WGM}
\tsc{QE}
\tsc{EP}
\tsc{PMS}
\tsc{BEC}
\tsc{DE}

\begin{document}
\let\WriteBookmarks\relax
\def\floatpagepagefraction{1}
\def\textpagefraction{.001}
\shorttitle{}
\shortauthors{Fuxin Zhang et~al.}

\title [mode = title]{Deep Learning Based Automatic Modulation Recognition: Models, Datasets, and Challenges}             

\author[1]{Fuxin~Zhang}

\author[1,2]{Chunbo~Luo}
\cormark[1]
\cortext[cor1]{Corresponding author}

\author[1]{Jialang~Xu}

\author[1]{Yang~Luo}

\author[3]{FuChun~Zheng}

\address[1]{School of Information and Communication Engineering, University of Electronic Science and Technology of China, Chengdu, China}
\address[2]{Department of Computer Science, University of Exeter, Exeter, EX4 4RN, UK}
\address[3]{School of Electronic and Information Engineering, Harbin Institute of Technology (Shenzhen), China}

\nonumnote{\textit{E-mail addresses}: c.luo@uestc.edu.cn (C. Luo), fxzhang@std.uestc.edu.cn (F. Zhang).}

\begin{abstract}
Automatic modulation recognition (AMR) detects the modulation scheme of the received signals for further signal processing without needing prior information, and provides the essential function when such information is missing. Recent breakthroughs in deep learning (DL) have laid the foundation for developing high-performance DL-AMR approaches for communications systems. Comparing with traditional modulation detection methods, DL-AMR approaches have achieved promising performance including high recognition accuracy and low false alarms due to the strong feature extraction and classification abilities of deep neural networks. Despite the promising potential, DL-AMR approaches also bring concerns to complexity and explainability, which affect the practical deployment in wireless communications systems. This paper aims to present a review of the current DL-AMR research, with a focus on appropriate DL models and benchmark datasets. We further provide comprehensive experiments to compare the state of the art models for single-input-single-output (SISO) systems from both accuracy and complexity perspectives, and propose to apply DL-AMR in the new multiple-input-multiple-output (MIMO) scenario with precoding. Finally, existing challenges and possible future research directions are discussed.
\end{abstract}

\begin{keywords}
Automatic modulation recognition \sep Deep learning \sep Neural networks  \sep Modulation
\end{keywords}
\maketitle

\section{Introduction}
\label{Introduction}
{A}{utomatic} modulation recognition (AMR) provides essential modulation information of the incoming radio signals, especially non-cooperative radio signals, and plays a key role in various scenarios including cognitive radio, spectrum sensing, signal surveillance, interference identification, etc \cite{9395503,dobre2007survey,9488834}. It aims to detect the modulation scheme of wireless communications signals automatically without prior information, and has attracted significant research interest in recent years \cite{o2016convolutional}. During the transmission process, signals generated by the transmitter are usually reshaped by adversarial factors within the radio frequency chain, such as noise, multi-path fading, shadow fading, center frequency offset, and sample rate offset \cite{o2016convolutional}. The structural characteristics of these signals may also be distorted due to poor hardware design or crystal oscillator drifting, causing difficulty to distinguish different modulation schemes. As an important step between signal detection and demodulation, AMR provides the essential function to detect modulation schemes. With the rapid development of wireless communications technologies, the modulation schemes of signals will become more complex and diverse to meet the needs of increasingly complex communications scenarios, so there is an urgent need to design effective AMR models that are robust in harsh radio environments.

Traditional research on AMR can be grouped into two categories: likelihood theory-based AMR (LB-AMR) \cite{dulek2017online,wei2000maximum,xu2010likelihood} and feature-based AMR (FB-AMR) \cite{hazza2013overview}. LB-AMR methods often receive optimal recognition accuracy in the sense of Bayesian estimation, but they have high computational complexity. FB-AMR methods mainly focus on learning representative features from training samples and classifying incoming signals using the trained models. Typical types of features used in FB-AMR include instantaneous time-domain features \cite{chan1985identification}, transform domain features \cite{hong1999identification}, and statistical features \cite{liu2006novel,swami2000hierarchical}. Machine learning models are increasingly employed for classification, including artificial neural network \cite{2004Automatic,xu2008digital}, decision tree \cite{yuan2004modulation}, and support vector machine \cite{park2008automatic}, etc. By contrast, FB-AMR methods usually yield sub-optimal solutions but with low computational complexity and multiple-modulation identification capability. 

\begin{figure}[htbp]
	\centering
	    \includegraphics[scale=0.65]{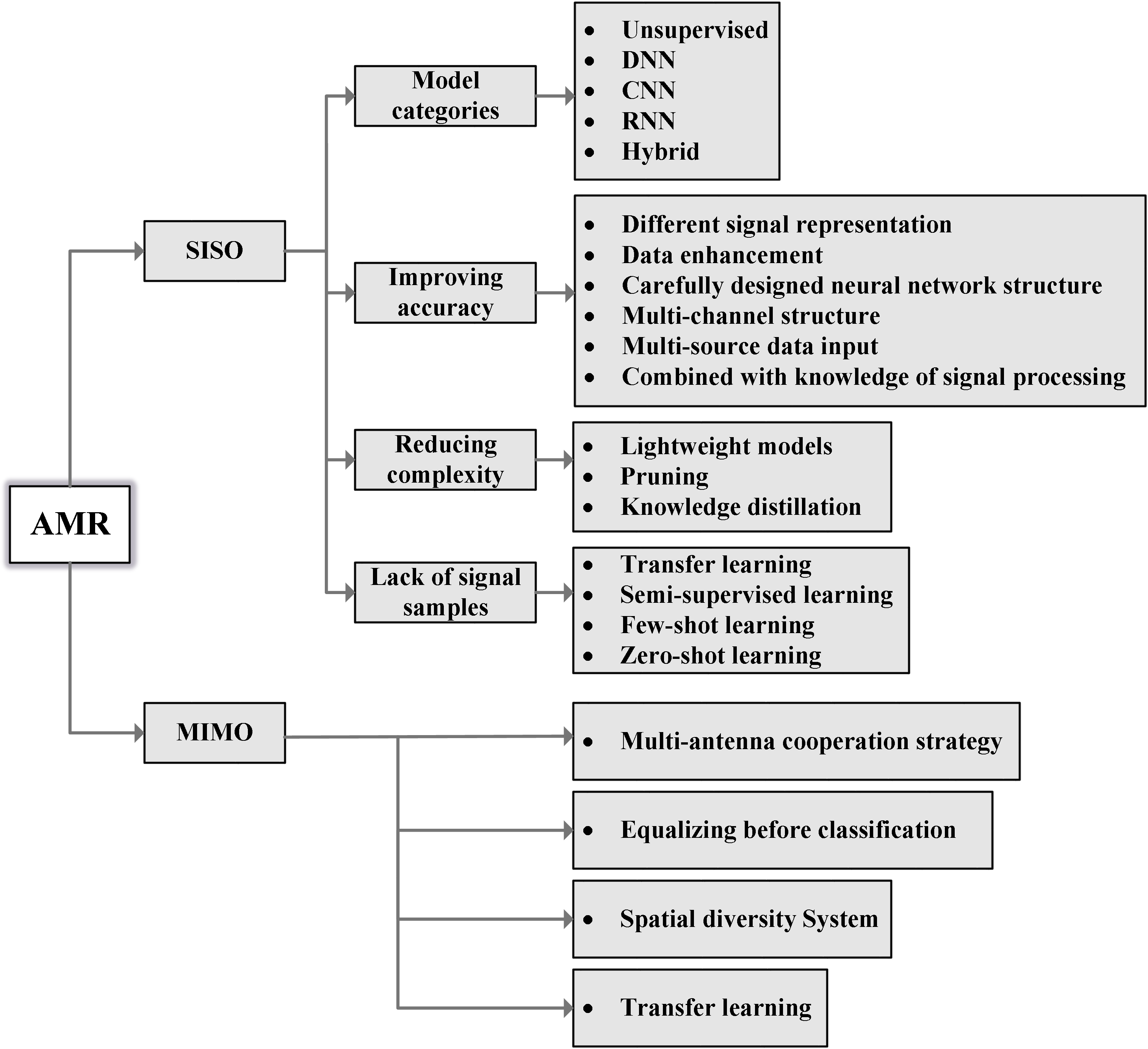}
	\caption{Taxonomy of existing DL-AMR research.}
	\label{FIG:1}
\end{figure}

Deep learning (DL) has achieved breakthroughs in a range of challenging applications where traditional methods struggle to achieve promising performance, such as natural language processing and image processing. The strong feature extraction capability empowered by stacked multi-layers of artificial neurons has inspired extended research in the conquest of modulation detection, where some pioneering DL based methods have been proposed, often outperforming traditional LB- and FB- AMR approaches \cite{o2016convolutional, rajendran2018deep}. Furthermore, existing research on DL-AMR mainly focuses on single-input-single-output (SISO) systems while there is increasing coverage on DL-AMR for multiple-input-multiple-output (MIMO) systems recently. DL-AMR for MIMO would provide new perspectives in this rapidly developing field. We present our taxonomy of existing DL-AMR research in Fig. \ref{FIG:1}.

This paper aims to provide a review of current DL-AMR approaches, outline the key challenges of this field, and shed light on future promising opportunities. We first introduce the general framework of DL-AMR and systematically analyze the existing DL-AMR models for SISO systems. The benchmark AMR datasets for SISO systems are summarized with a focus on their key characteristics. We further provide an overview of DL-AMR for MIMO systems, and then propose to apply AMR in the MIMO communications system containing precoding. In addition, extensive experiments with the main state-of-the-art DL-AMR models showed detailed comparisons for practitioners to assess the different models according to the applications (source code can be found at https://github.com/Richardzhangxx after publication). Finally, we discuss the remaining challenges of AMR and future research opportunities. 

Note that this paper is conducted in parallel and independent from the recently published survey papers (e.g., \cite{9576081} and \cite{9399081}). We analyze the models, datasets, and challenges for AMR from different perspectives including the latest proposed models, and experimentally compare the most representative and state-of-the-art models in AMR field under a unified platform with the same parameter settings, providing a comprehensive reference for future research on AMR. The feasibility of applying DL-AMR in the MIMO system with precoding is also experimentally verified, showing that SISO's method can be transferred to MIMO systems.

\section{DL models for AMR in SISO systems}
After the signal passes through the channel and is sampled, the equivalent baseband signal can be expressed by:
\begin{equation}
y[l]=A[l]e^{j(\omega l+\varphi)} x[l]+n[l], l=1, \ldots, L,
\end{equation}
where $x[l]$ is the signal modulated by the transmitter in a certain modulation scheme, $n[l]$ denotes the complex Additive Gaussian Noise (AWGN), $A[l]$ represents the channel gain, $\omega$ is the frequency offset, $\varphi$ is the phase offset, $y[l]$ denotes the $l$-th value observed by the receiver and $L$ is the number of symbols in a signal sample. To facilitate data processing and modulation recognition, the received signals can be stored in in-phase/quadrature (I/Q) form, denoted as $\textbf{y} = [\Re\{y[1]\}, ..., \Re\{y[L]\}; \Im\{y[1]\}, ..., \Im\{y[L]\}]$. Modulation recognition needs to identify the modulation scheme of the transmitted signal $x[l]$ from $y[l]$, while the structural characteristics of the signals may have been distorted.
\begin{figure}[htbp]
	\centering
	    \includegraphics[scale=0.38]{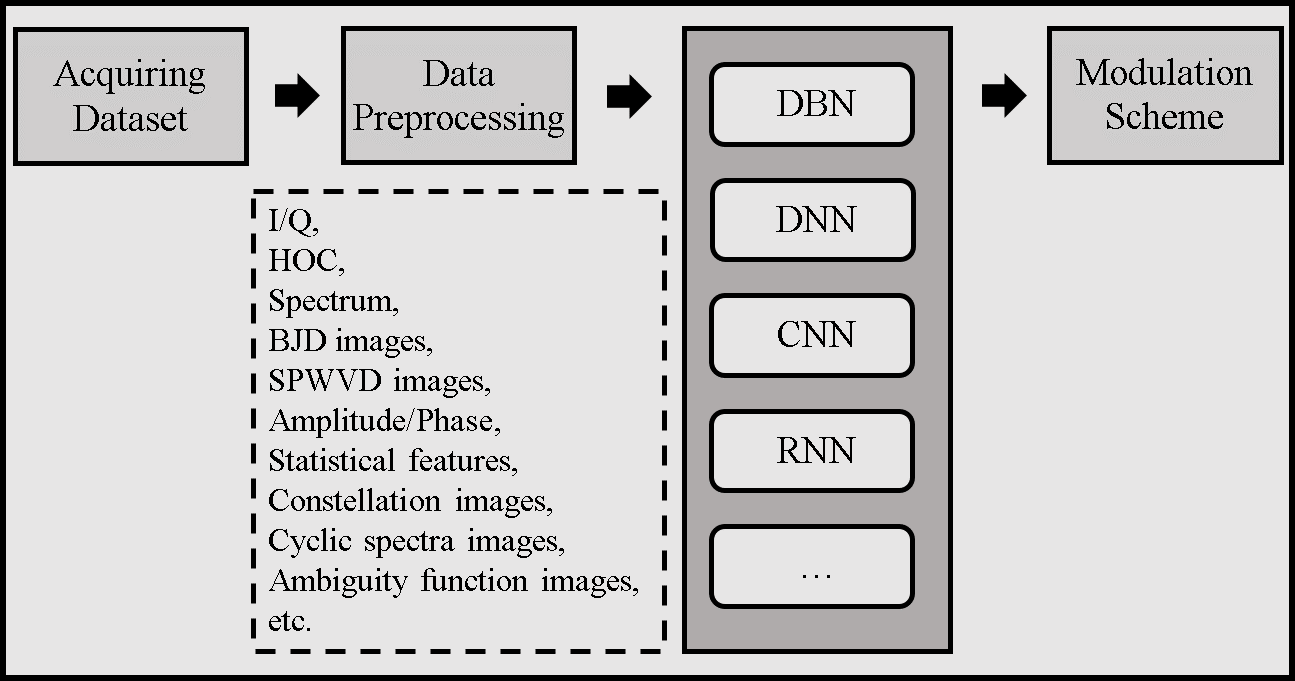}
	\caption{The general structure of DL-AMR.}
	\centering
	\label{FIG:2}
\end{figure}

\begin{figure}[htbp]
    \centering
        \includegraphics[scale=0.038]{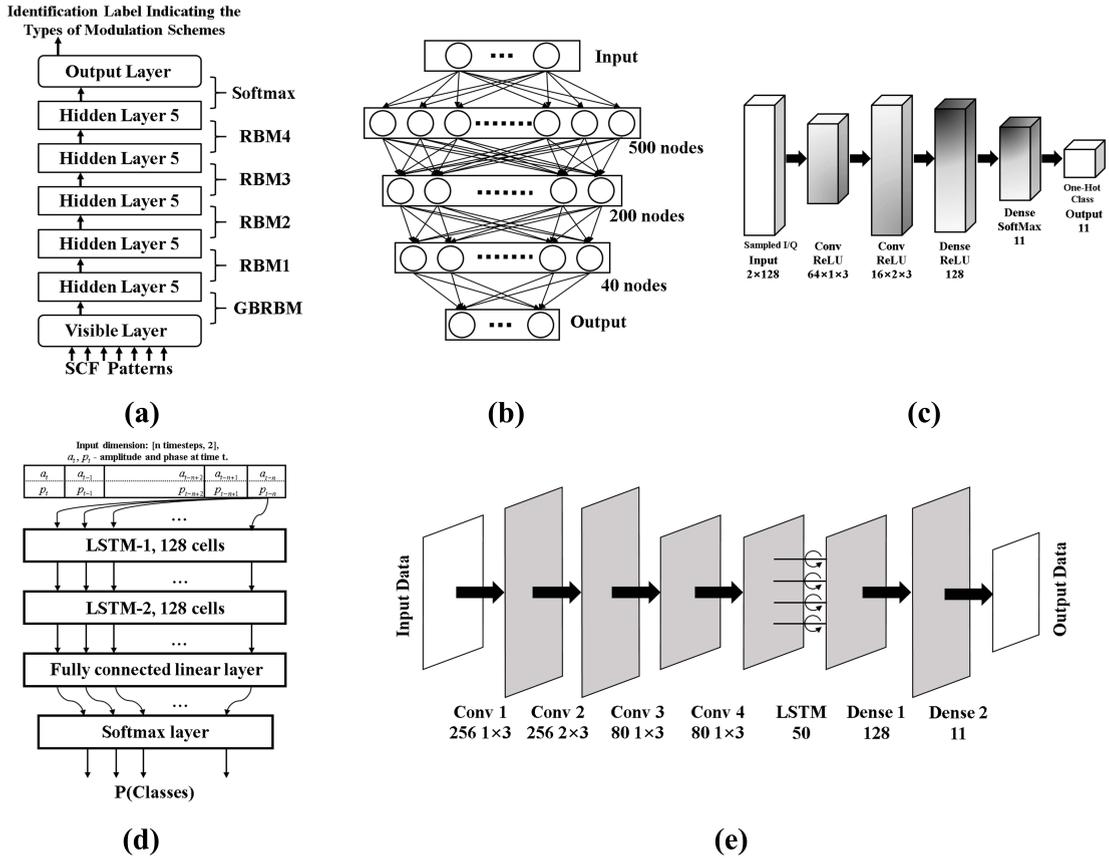}
    \caption{Representative DL-AMR models for each neural network type. (a) Unsupervised \cite{mendis2019deep}, (b) DNN \cite{2017Deep}, (c) CNN \cite{o2016convolutional}, (d) RNN \cite{rajendran2018deep}, (e) Hybird \cite{liu2017deep}.}
    \label{FIG:3}
\end{figure}

DL-AMR generally contains three steps, including preprocessing of the received signals, feature extraction, and modulation classification, as shown in Fig. \ref{FIG:2}. Preprocessing is used to adapt the data format of the signals for subsequent feature extraction and training. Feature extraction and feature classification can be completed by a deep neural network in an end-to-end fashion or divided into two steps, where features are extracted first by traditional methods in the preprocessing part and then used to train a classification model. The recent representative works on SISO-AMR using deep learning are summarized in Table \ref{tbl1}, and the representative DL-AMR model for each neural network type is shown in Fig \ref{FIG:3}.

\begin{table}[htbp]
\caption{List of deep learning based SISO-AMR models.}
\label{tbl1}
\resizebox{\textwidth}{!}{
\begin{tabular}{|cccccc|}
\hline
\textbf{Model type} & \textbf{Model} & \textbf{Author} & \textbf{Dataset} & \textbf{Input} & \textbf{Main structure} \\ \hline
\multicolumn{1}{|c|}{} & SCF-DBN & Mendis et al. \cite{mendis2016deep,mendis2019deep} & Non-public dataset & Spectral correlation function & Restricted boltzmann machines \\ \cline{2-6} 
\multicolumn{1}{|c|}{\textbf{Unsupervised}}  & Autoencoder1 & Afan et al. \cite{SpaseAli} & Non-public dataset & 3 fourth-order cumulants & 3 fully connected layers \\ \cline{2-6} 
\multicolumn{1}{|c|}{} & Autoencoder2 & Afan et al. \cite{ali2017k} & Non-public dataset  & I/Q & 2 fully connected layers\\ \cline{2-6}
\multicolumn{1}{|c|}{} & Autoencoder3 & Dai et al. \cite{AF-SAE} & Non-public dataset &  Ambiguity function (AF) images & Two sparse auto-encoders (SAE) \\ \hline 
\multicolumn{1}{|c|}{} & DNN1 & Lee et al. \cite{2017Deep} & Non-public dataset & Statistical features & 4 fully connected layers \\ \cline{2-6} 
\multicolumn{1}{|c|}{\textbf{DNN}} & DNN2 &  Xie et al. \cite{2019Deepdnn} & Non-public dataset  & 6 high-order cumulants & 4 fully connected layers\\ \cline{2-6} 
\multicolumn{1}{|c|}{} & DNN3 &  Shi et al. \cite{2019Particle} & Non-public dataset  & 12 signal features & PSO-DNN \\ \hline
\multicolumn{1}{|c|}{} & CNN1 & O’Shea et al. \cite{o2016convolutional} & RML & I/Q & CNN \\ \cline{2-6} 
\multicolumn{1}{|c|}{} & CNN2 & Tekbıyık et al. \cite{tekbiyik2020robust} & HisarMod2019.1 & I/Q & CNN \\ \cline{2-6} 
\multicolumn{1}{|c|}{} & CNN3 & Kumar et al. \cite{CNN3} & Non-public dataset & Constellation density matrix (CDM) & ResNet-50 and Inception ResNet V2  \\ \cline{2-6} 
\multicolumn{1}{|c|}{} & CFCN & Huang et al. \cite{CFCN} & Non-public dataset &  Grid constellation matrix (GCM) & CNN with contrastive loss  \\ \cline{2-6} 
\multicolumn{1}{|c|}{} & CCNN & Huang et al. \cite{CCNN} & Non-public dataset & Two kinds of constellation diagram & CNN with compressive loss  \\ \cline{2-6} 
\multicolumn{1}{|c|}{} & CNN4 & Ma et al. \cite{Multifeafusion} & Non-public dataset & Cyclic spectra image and constellation diagram & Two-branch CNN \\ \cline{2-6}
\multicolumn{1}{|c|}{} & CNN5 & Hiremath et al. \cite{hiremath2019deep} & Non-public dataset & I/Q and DOST& Two-branch CNN \\ \cline{2-6}
\multicolumn{1}{|c|}{} & CNN6 & Zhang et al. \cite{BJD} & RML & SPWVD images, BJD images, and hand-crafted features & Two-branch CNN \\ \cline{2-6}
\multicolumn{1}{|c|}{} & CNN7 & Wang et al. \cite{eyediagram} & Non-public dataset & Eye-diagram & CNN \\ \cline{2-6}
\multicolumn{1}{|c|}{\textbf{CNN}} & CNN8 & Lee et al. \cite{FPimages} & Non-public dataset & Feature point (FP) image & CNN \\ \cline{2-6}
\multicolumn{1}{|c|}{} & CNN9 & Shi et al. \cite{shi2022combining} & RML & I/Q & Squeeze-and-excitation (SE) block \\ \cline{2-6}
\multicolumn{1}{|c|}{} & AlexNet & Peng et al. \cite{2018Modulation} & Non-public dataset & 3-channel constellation images & AlexNet\\ \cline{2-6} 
\multicolumn{1}{|c|}{} & GoogLeNet & Peng et al. \cite{2018Modulation} & Non-public dataset & 3-channel constellation images & GoogLeNet \\ \cline{2-6} 
\multicolumn{1}{|c|}{} & MCNET & Thien et al. \cite{huynh2020mcnet} & RML & I/Q & CNN \\ \cline{2-6} 
\multicolumn{1}{|c|}{} & TRNN & Zhang et al. \cite{9467343} & Publicly available dataset & I/Q & CNN \\ \cline{2-6} 
\multicolumn{1}{|c|}{} & DrCNN & Wang et al. \cite{wang2019data} & Non-public dataset & I/Q + Constellation images & Two CNN  \\ \cline{2-6}
\multicolumn{1}{|c|}{} & IC-AMCNet & Hermawan et al. \cite{hermawan2020cnn} & RML & I/Q & \begin{tabular}[c]{@{}c@{}}CNN + Gaussian noise\end{tabular} \\ \cline{2-6} 
\multicolumn{1}{|c|}{} & ResNet & Liu et al. \cite{liu2017deep} & RML & I/Q & ResNet \\ \cline{2-6} 
\multicolumn{1}{|c|}{} & DenseNet & Liu et al. \cite{liu2017deep} & RML & I/Q & DenseNet \\ \cline{2-6} 
\multicolumn{1}{|c|}{} & CM + CNN & Yashashwi et al. \cite{yashashwi2018learnable} & RML & I/Q & Correction module + CNN \\ \cline{2-6} 
\multicolumn{1}{|c|}{} & SigNet & Chen et al. \cite{9580446} & RML and Non-public dataset & Square matrix & Signal-to-matrix (S2M) + Trainable filters + ResNet50 \\ \cline{2-6} 
\multicolumn{1}{|c|}{} & SCNN & Zeng et al. \cite{zeng2019spectrum} & RML & Spectrum & CNN \\  \cline{2-6} 
\multicolumn{1}{|c|}{} & Bispectrum-Alexnet & Li et al. \cite{Bispectrum-Alexnet} & Non-public dataset & Amplitude spectrums of bispectrum (ASB)  & Alexnet \\  \cline{2-6} 
\multicolumn{1}{|c|}{} & CCES-ResNet & Ma et al. \cite{CCESRESTNET} & Non-public dataset & Cyclic correntropy spectrum (CCES) & ResNet \\ \hline
\multicolumn{1}{|c|}{} & GRU & Hong et al. \cite{hong2017automatic} & RML & I/Q & 2 GRU layers \\ \cline{2-6} 
\multicolumn{1}{|c|}{\textbf{RNN}} & LSTM & Rajendran et al. \cite{rajendran2018deep} & RML & Amplitude/Phase & 2 LSTM layers \\ \cline{2-6} 
\multicolumn{1}{|c|}{} & LSTM-DNN & Hu et al. \cite{ShishengHu} & Non-public dataset & I/Q & LSTM + Temporal attention + DNN \\  \cline{2-6}
\multicolumn{1}{|c|}{} & DAE & Ke et al. \cite{9487492} & RML & Amplitude/Phase & LSTM +  DNN \\  \hline
\multicolumn{1}{|c|}{} & CLDNN & West et al. \cite{west2017deep} & RML & I/Q & \begin{tabular}[c]{@{}c@{}}CNN + LSTM + Skip connections\end{tabular} \\ \cline{2-6}
\multicolumn{1}{|c|}{} & CLDNN2 & Liu et al. \cite{liu2017deep} & RML & I/Q & CNN + LSTM \\ \cline{2-6} 
\multicolumn{1}{|c|}{} & CGDNet & Njoku et al. \cite{9349627} & RML & I/Q & CNN + GRU +DNN \\ \cline{2-6} 
\multicolumn{1}{|c|}{} & PET-CGGDNN & Zhang et al. \cite{zhang2021efficient} & RML & I/Q & CNN + GRU +DNN \\ \cline{2-6}
\multicolumn{1}{|c|}{\textbf{Hybrid}} & MCLDNN & Xu et al. \cite{xu2020spatiotemporal} & RML & I/Q, I and Q & \begin{tabular}[c]{@{}c@{}}CNN + LSTM + Multi-channel\end{tabular} \\ \cline{2-6}
\multicolumn{1}{|c|}{} & DBN + SNN & Ghasemzadeh et al. \cite{9188007} & RML & I/Q + Amplitude/Phase + High-order statistical feature & DBN + SNN \\ \cline{2-6} 
\multicolumn{1}{|c|}{} & GS-QRNN & Ghasemzadeh et al. \cite{9672088} & RML & I/Q & GRU + CNN \\ \cline{2-6} 
\multicolumn{1}{|c|}{} & CNN-LSTM-IQFOC & Zhang et al. \cite{IQFOC} & RML & I/Q + Fourth order cumulants (FOC) & CNN + LSTM \\  \cline{2-6}
\multicolumn{1}{|c|}{} & CNN-LSTM & Zhang et al. \cite{zhang2020automatic} & RML & I/Q + Amplitude/Phase & CNN + LSTM \\  \cline{2-6}
\multicolumn{1}{|c|}{} & MLDNN & Chang et al. \cite{9462447} & RML & I/Q + Amplitude/Phase & CNN + BiGRU + SAFN \\  \cline{2-6}
\multicolumn{1}{|c|}{} & MCF & Wang et al. \cite{9427151} & RML & I/Q + Amplitude/Phase + Constellation diagram & CNN + IndRNN \\  \hline
\end{tabular}}
\end{table}

\subsection{Unsupervised models}
Unsupervised models like deep belief networks (DBN) and autoencoder have been applied for AMR. DBN models consist of multiple layers of restricted Boltzmann machines (RBM), which represent a probabilistic generation method. The authors of \cite{mendis2016deep,mendis2019deep} integrate spectral correlation function (SCF) and DBN to learn complex patterns effectively. Furthermore, they extend their preliminary work by implementing it and analyzing the reduction of logic utilization on FPGA, further demonstrating the effectiveness of this method. DBN has a greedy layer-by-layer unsupervised training procedure, which is too complex to be effectively applied to large-scale learning problems. Due to the high computational complexity of DBN and the development of other deep learning methods, DBN models are gradually losing popularity in AMR. In \cite{ali2017k}, a low complexity classifier has been proposed to use the $k$-sparse autoencoder to extract features of signals and then classify their modulation schemes with the features extracted. The overall complexity has been reduced by using only $k$ largest hidden nodes to reconstruct the input rather than using all the hidden nodes, while keeping the classification accuracy to a near-optimal level. The author further proposes a nonnegativity constraint training algorithm in \cite{SpaseAli} to minimize the reconstruction error, which takes fourth-order cumulants as input instead of in-phase/quadrature (I/Q) data in \cite{ali2017k}. Another form of input called ambiguity function (AF) is adopted in \cite{AF-SAE}, which inputs 28 × 28 AF images into sparse auto-encoders (SAE) for modulation recognition. However, most unsupervised deep learning methods have only been tested to classify a few modulation schemes, and their classification performance needs to be generalized to more practical applications. 

\subsection{DNN-based models}
The deep neural network (DNN) model has the structure of a fully connected feedforward network with several hidden layers. Due to the capability of DNN to extract the complex feature in high dimensional space, a four-layer DNN model is adopted in \cite{2017Deep} to classify the modulation schemes from various statistical features, e.g., high-order cumulants, kurtosis, and skewness. Different modulation recognition schemes are considered in \cite{2019Deepdnn}, where six high-order cumulants are used as input and a similar DNN structure is used for modulation recognition. To avoid selecting the number of neural network nodes manually, \cite{2019Particle} introduces the particle swarm optimization (PSO) scheme to optimize the number of hidden layer nodes, which can improve recognition accuracy compared to traditional DNN.

\subsection{CNN-based models}
Convolutional neural network (CNN) models have shown exceptional capability in processing data that exhibit spatial characteristics, e.g., image segmentation, object detection and classification. Researches on AMR have introduced CNNs to detect the signal modulation schemes by utilizing their spatial feature extraction power. Depending on the input data types, existing CNN-based AMR approaches can be grouped into two categories: CNN models with raw I/Q inputs and CNN models with pre-processed inputs. Another category is efficient CNN architecture design, which is to meet the latency and complexity requirements in advanced communications systems. These categories are analyzed below.

\textbf{\emph{1) Typical CNN models with raw I/Q data inputs}: }
By using a simple four-layer CNN model and taking I/Q data as inputs, the  potential of CNN applied to AMR is firstly explored in \cite{o2016convolutional} which achieves higher recognition accuracy than many traditional methods. Authors of \cite{tekbiyik2020robust} further propose updated CNN models by adapting the number of layers or hyper-parameters. Because these simple CNN model structures may not fully extract representative features from the I/Q input signals, a few improved CNN models that utilize a mixture of complex layers or transformation operations have since been applied in AMR. For example, inspired by the winner architecture of the ImageNet 2015 competition, new DL-AMR models based on residual neural network (ResNet) and densely connected network (DenseNet) \cite{liu2017deep} are proposed to allow features learned from multiple layers to be effectively transmitted to the detection module. A real-time signal modulation classification model is presented in \cite{9467343}, which identifies OFDM modulation signals by a triple-skip residual neural network (TRNN). However, their improvements in recognition accuracy come with the significantly increased cost of computational complexity. Since many CNN-based AMR models directly adopt the methods proposed for computer vision tasks, they neglect the intrinsic signal characteristics in communications systems. To address such limitations, the authors of \cite{yashashwi2018learnable} propose to estimate the carrier frequency offsets and phase noise of the received signals before feeding the data into CNN models, by applying a trainable function that calculates the phase and frequency offsets (correction parameters) from the received signals which are corrected by reducing the offsets.

\textbf{\emph{2) Typical CNN models with pre-processed inputs}:}
CNN models taking inputs of raw I/Q signals, which are the results of multiple processing modules through the RF chain coupled with channel effects, if fed into the model directly, would miss important characteristics that could be easily extracted by classical signal processing methods, e.g., high-order cumulants, spectrum images, and constellation diagrams, etc. Combining traditional FB-AMR methods with CNN models has shown significant potential to overcome such limitations. For example, a short-time discrete Fourier transform is used in \cite{zeng2019spectrum} to transform one-dimensional radio signals into spectrum images, and then a Gaussian filter is added to reduce noise. The proposed SCNN model has achieved better recognition accuracy than the benchmark CNN1 \cite{o2016convolutional} models, demonstrating the effectiveness of this strategy. The Bispectrum-Alexnet method proposed in \cite{Bispectrum-Alexnet} feeds the CNN with amplitude spectrums of bispectrum (ASB) since bispectrum can suppress the effect of white noise. Another spectrum called cyclic correntropy spectrum (CCES) is introduced in \cite{CCESRESTNET} to improve the recognition performance under non-Gaussian noise, and ResNet is then used for classification. In \cite{hiremath2019deep}, I/Q and the discrete orthonormal Stockwell transform (DOST) of I/Q sequences are combined to represent the received signal for CNN-based modulation recognition. Moreover, constellation diagram is another widely used signal representation method for AMR, and the features could be extracted directly from the constellation diagram to determine the modulation scheme. Constellation diagram is converted into 3-channel image in \cite{2018Modulation} to leverage the colored images processing ability of CNN models, which is further classified by AlexNet and GoogLeNet models. Other representative CNN models, e.g., ResNet50 and Inception ResNet V2 are also applied in \cite{CNN3}, which transfer constellation points to constellation density matrix (CDM) and mask the CDM with a proper filter to remove information corrupted by high noise. Note that the confusion between 16QAM and 64QAM is very severe, \cite{wang2019data} uses two CNN models to improve the recognition accuracy, and trains two models by using I/Q data and constellation diagrams respectively. The CNN based on constellation diagrams is mainly used to solve the problem of confusion between 16QAM and 64QAM. The cross-entropy loss function is the most commonly used loss function in AMR, perhaps it is not the most suitable loss function for all DL-AMR models. On this basis, the authors propose to use contrastive loss \cite{CFCN} and compressive loss \cite{CCNN} as the loss function to enhance the discrepancy of the extracted representations among different modulations, and the corresponding models utilize grid constellation matrix (GCM) and multiple constellation images as input, respectively. To leverage the strong noise resistibility of cyclic spectra image and high-order modulation recognition capability brought by constellation diagram, a two-branch CNN is built in \cite{Multifeafusion} to extract multi-features and then fuse them together. Also based on the idea of feature fusion, \cite{BJD} applied a multimodality fusion model, which fuses different image features extracted by CNNs and eight handcrafted features, to obtain more discriminating features. Some other input signal representations, such as eye diagrams \cite{eyediagram}, feature point (FP) images \cite{FPimages}, and square feature matrix \cite{9580446} are also used as input to the DL-AMR model.

\textbf{\emph{3) Efficient CNN architecture design}:}
In order to reduce the projected latency in beyond fifth-generation (B5G) communications systems, an improved CNN-based AMR model \cite{hermawan2020cnn} is developed by adopting a small number of filters in each layer and reducing the overall number of trainable parameters. The processing time of the model can be less than 0.01ms to comply with the B5G communications requirement. Despite the reduced number of trainable parameters in this model, it can still maintain a high recognition accuracy. To meet the baseline requirements of 5G services in terms of ultra-reliability and low-latency, a cost-efficient and high-performed CNN is proposed in \cite{huynh2020mcnet} to effectively classify the modulations schemes, which adopts various asymmetric kernels in parallel in each convolutional block and skip connections in a block-wise manner throughout the network. Another more lightweight model is proposed in \cite{shi2022combining}, which introduces the channel attention mechanism by employing the squeeze-and-excitation (SE) block.

\subsection{RNN-based models}
In addition to the spatial correlation characteristics, wireless communications signals also carry temporal correlation features that can be learned by machine learning models such as recurrent neural network (RNN) for modulation recognition. A few novel model architectures based on RNN have been proposed recently and obtained state-of-the-art performance in AMR. For example, a new AMR method based on RNN is proposed in \cite{hong2017automatic} which uses gated recurrent units (GRU) to obtain better recognition accuracy than a few CNN models. The author of \cite{rajendran2018deep} transforms the I/Q signals into amplitude and phase, and inputs them into an LSTM which also achieves high recognition accuracy. The models using only two RNN layers \cite{hong2017automatic,rajendran2018deep} have already demonstrated outstanding accuracy performance in AMR tasks, confirming the distinct temporal features of communications signals and the strong feature extraction ability of RNNs. Another LSTM-based model proposed in \cite{ShishengHu}, which incorporates fully connected layers and a temporal attention mechanism, is proved to be robust to uncertain noise conditions. An LSTM denoising auto-encoder is designed in \cite{9487492} to automatically infer modulation schemes using a compact RNN architecture readily implemented on a low-cost computational platform while exceeding state-of-the-art accuracy in \cite{rajendran2018deep}. Benefitting from the respective advantages of RNN and CNN, researchers have further proposed hybrid models combining RNN and CNN to further improve the AMR performance, as shown below. 

\subsection{Hybrid models}
Since pure CNN or RNN models focus on either the spatial features or temporal features of AMR signals, only using one type of them may not achieve optimal performance. Therefore, researchers have proposed to combine the characteristics of both types of neural network layers to build hybrid models for AMR. A convolutional long-short-term deep neural networks (CLDNN) model has been proposed in \cite{west2017deep} consisting of one LSTM and three CNN layers. This model has a skip connection before LSTM that bypasses two CNN layers, which are intended to provide a longer time context for the extracted features. The output of the first convolution layer is concatenated with the output of the third convolution layer before feeding into the LSTM and then the LSTM layer can extract more effective temporal features which can yield better recognition accuracy and more stable gradient descent than other architectures such as CNN1 \cite{o2016convolutional}. Another CLDNN model named CLDNN2 with no bypass layer connections is shown in \cite{liu2017deep}, which can achieve higher recognition accuracy than CLDNN \cite{west2017deep} at the cost of increasing the number of layers and parameters. To reduce the computational complexity, \cite{9672088} propose a stacking quasi-recurrent neural network (SQRNN) to mimic recurrent layer operations, which is implemented by low-complexity convolutional layers. Another cost-efficient hybrid neural network composed of CNN, GRU, and DNN is introduced in \cite{9349627}. Besides, an efficient DL-AMR model based on phase parameter estimation and transformation is proposed in \cite{zhang2021efficient}, which achieves high recognition accuracy but with the least parameters. Inspired by the excellent feature extraction characteristics of hybrid models and the complementary information existing in separate channels, a new multi-channel deep learning model is proposed to extract features using single and combined I/Q symbols of received data and from both temporal and spatial perspectives \cite{xu2020spatiotemporal}, achieving the current benchmark performance in AMR. To enrich the input data with features that can be easily extracted, \cite{IQFOC} combines raw IQ data and fourth order cumulants (FOC) as input. Additionally, a CNN-LSTM based dual-stream structure is proposed in \cite{zhang2020automatic}, which takes amplitude/phase (A/P) and I/Q data as the input of the two streams. Another study to effectively fuse I/Q and A/P is presented in \cite{9462447}, where an innovative step attention fusion network (SAFN) block is proposed to merge all step outputs from the bidirectional gated recurrent unit (BiGRU) layer by different weights. In addition to I/Q and A/P, higher order statistical features are used as additional features, using elementwise add operation to fuse these features \cite{9188007}. A multi-cue fusion (MCF) network based on CNN and independently recurrent neural network(IndRNN) is proposed in \cite{9427151}, which fuses a signal cue and a visual cue to jointly determine the modulation scheme. While such models achieve high recognition accuracy, the multi-input structure is more expensive than the models with a single channel. With the advancement of deep learning, more hybrid models and signal representation methods could be proposed to further push the boundary in this field.

\section{AMR open source datasets for SISO systems}
\label{section:AMR open source datasets}

\begin{table}[htbp]
\caption{Main AMR open datasets for SISO systems.}
\label{tbl2}
\resizebox{\textwidth}{!}{
\begin{tabular}{|cccccc|}
\hline
\textbf{\begin{tabular}[c]{@{}c@{}}Dataset Name\end{tabular}} & \textbf{\begin{tabular}[c]{@{}c@{}}Modulation Schemes\end{tabular}} & \textbf{\begin{tabular}[c]{@{}c@{}}Sample Dimension\end{tabular}} & \textbf{\begin{tabular}[c]{@{}c@{}}Dataset Size\end{tabular}} & \textbf{\begin{tabular}[c]{@{}c@{}}SNR Range(dB)\end{tabular}} & 
\textbf{Characteristics} \\ \hline
\textit{\begin{tabular}[c]{@{}c@{}}RML\\  2016.04c\end{tabular}} & \begin{tabular}[c]{@{}c@{}}11 classes (8PSK, BPSK, CPFSK, GFSK, PAM4, \\ AM-DSB, AM-SSB, 16QAM, 64QAM, QPSK, WBFM)\end{tabular} & 2*128 & 162060 & -20:2:18 & \begin{tabular}[c]{@{}c@{}}The original version of  \\ the AMR dataset, relatively small.\end{tabular} \\ \hline
\textit{\begin{tabular}[c]{@{}c@{}}RML\\  2016.10a\end{tabular}} & \begin{tabular}[c]{@{}c@{}}11 classes (8PSK, BPSK, CPFSK, GFSK, PAM4, \\ 16QAM, AM-DSB, AM-SSB, 64QAM, QPSK, WBFM)\end{tabular} & 2*128 & 220000 & -20:2:18 & \begin{tabular}[c]{@{}c@{}}A cleaner and more standardized \\ version of RML2016.04c.\end{tabular} \\ \hline
\textit{\begin{tabular}[c]{@{}c@{}}RML\\  2016.10b\end{tabular}} & \begin{tabular}[c]{@{}c@{}}10 classes (8PSK, BPSK, CPFSK, GFSK, PAM4, \\ AM-DSB, 16QAM, 64QAM, QPSK, WBFM)\end{tabular} & 2*128 & 1200000 & -20:2:18 & \begin{tabular}[c]{@{}c@{}}A larger version of RML2016.10a,\\ but not including AM-SSB.\end{tabular} \\ \hline
\textit{\begin{tabular}[c]{@{}c@{}}RML\\  2018.01a\end{tabular}} & \begin{tabular}[c]{@{}c@{}}24 classes (OOK, 4ASK, 8ASK, BPSK, QPSK,\\  8PSK, 16PSK, 32PSK, 16APSK, 32APSK, 64APSK,\\ 128APSK, 16QAM, 32QAM, 64QAM, 128QAM, \\ 256QAM, AM-SSB-WC, AM-SSB-SC, AM-DSB-WC,\\ AM-DSB-SC, FM, GMASK, OQPSK)\end{tabular} & 2*1024 & 2555904 & -20:2:30 & \begin{tabular}[c]{@{}c@{}}This dataset is very large and contains \\ more kinds of modulation schemes.\end{tabular} \\ \hline
\textit{HisarMod2019.1} & \begin{tabular}[c]{@{}c@{}}26 classes   (AM-DSB, AM-SC, AM-USB, AM-LSB,\\ FM, PM, 2FSK, 4FSK, 8FSK, 16FSK,  4PAM, \\ 8PAM, 16PAM, BPSK, QPSK, 8PSK, 16PSK,\\ 32PSK, 64PSK, 4QAM, 8QAM, 16QAM, \\ 32QAM, 64QAM, 128QAM, 256QAM)\end{tabular} & 2*1024 & 780000 & -20:2:18 & \begin{tabular}[c]{@{}c@{}}Contains signals from five different \\ wireless communications channels.\end{tabular} \\ \hline
\end{tabular}}
\end{table}

Learning-based AMR approaches heavily depend on datasets. Comprehensive high-quality datasets underpin the key aspects of DL-AMR, including model training, testing, and evaluation. 

Table \ref{tbl2} summarizes the main AMR datasets and compares their properties. The RML datasets \footnote{http://radioml.com} are generated using GNU radio by O’Shea et al. \cite{o2016convolutional,o2018over}, and RML2016.10a has been used in most research as a benchmark dataset. To simulate the data characteristics of modulated signals in real-world situations, the RML datasets take into account the common time-varying random channel effects in most wireless systems, which include center frequency offset, sample rate offset, additive white Gaussian noise, multi-path, and fading. RML2016.10a and RML2016.10b are generated by simulating the propagation properties in a harsh environment while RML2018.01A is produced in a relatively good real laboratory environment. To introduce a more comprehensive dataset, \cite{tekbiyik2020robust} create a new dataset called HisarMod2019.1 \footnote{http://dx.doi.org/10.21227/8k12-2g70} using simulation software Matlab with 26 modulation classes of signals passing through the channels. This dataset provides wireless signals under ideal, static, Rayleigh, Rician(k = 3), and Nakagami–m (m = 2) channel conditions with various numbers of channel taps. 

At present, most AMR research use RML2016.10a dataset, which has moderate data volume and contains some common analog and digital modulation types like QAM, PSK, FSK, and AM. The later RML2016.10b dataset has a larger data volume and requires more computing resources. The modulation schemes in RML2018.01a dataset increase to 24 and the data length has also increased from 128 to 1024, which requires very high computing resources. However, the rich data of RML2018.01a enable further research to develop more advanced models. Although HisarMod2019.1 has 26 modulation types, the dataset is generated in a better transmission environment and thus the modulation schemes can be more easily identified.

Generally, datasets for deep learning methods are difficult to collect, and demand significant resource investment. Some of the datasets just simulate the channels and signals rather than collect data from real wireless communications environments. Despite the clear advantages such as generating a large number of data samples in a short time, the real-world channel conditions are usually more complex and dynamic. To design a DL-AMR model that can work under real channel conditions would require training or fine-tuning on datasets that are obtained from real-world scenarios.

\section{DL models for AMR in MIMO systems}
\subsection{Introduction of the MIMO DL-AMR systems}
The received signal after passing through the MIMO channel can be given as:
\begin{equation}
\textbf{R}(n)=\textbf{H}\textbf{T}(n)+\textbf{G}(n), n=1, \ldots, N,
\end{equation}

where \textbf{H} is the MIMO channel matrix, \textbf{R}(n) is the received signal vector at $n$-th sample time, \textbf{T}(n) is the transmitted signal vector and \textbf{G}(n) is the additive white Gaussian noise (AWGN).

Different from SISO systems, MIMO systems can use the signals received by multiple antennas to jointly determine the modulation scheme. The authors of \cite{wang2020deep} proposed a cooperative DL-AMR method for MIMO systems, which obtains recognition sub-results of the signals received by each antenna through CNN, and then determines the final modulation schemes from the sub-results based on the voting and averaging strategy. To reduce the adverse effects caused by the channel, a CNN based zero-forcing (ZF) equalization AMR method is further proposed for MIMO systems \cite{wang2020automatic}. The input data of the CNN-based model is the signal pre-processed by a ZF equalizer. A more complex system considering space-time block-code (STBC) is proposed in \cite{STBCMIMO}, using sparse autoencoders(SAE) based DNN and radial basis 
function network for modulation recognition. Considering the generally few labeled samples and large unlabeled samples in realistic communication scenarios, a transfer learning-based semi-supervised AMR method for a MIMO system is further proposed in \cite{wang2020transfer}. Knowledge is transferred from the encoder layer of the reconstruction convolutional auto-encoder (CAE) to the feature layer of the AMR CNN by sharing their weights, where the reconstruction convolutional auto-encoder is trained by unlabeled samples and the AMR CNN is trained by labeled samples. Currently, only a few research has paid attention to applying deep neural networks for modulation recognition in MIMO systems, which is a very promising direction for developing novel methods based on the abundant and effective DL-AMR works in SISO systems.

\begin{table}[htbp]
\begin{center}
\caption{Representative deep learning models for AMR in MIMO systems.}
\label{tbl3}
\begin{tabular}{|c|c|c|c|}
\hline
\textbf{Model}      & \textbf{Author}       & \textbf{Input}                                & \textbf{Main structure}                     \\ \hline
Co-AMC     & Wang et al. \cite{wang2020deep}  & I/Q data of multiple antennas        & CNN and cooperative decision rules \\ \hline
CNN/ZF-AMC & Wang et al. \cite{wang2020automatic}  & The equalized I/Q sequence           & CNN                                \\ \hline
STBC-AMC   & Shah et al. \cite{STBCMIMO}  & Instantaneous and HOC-based features & SAE based DNN and RBFN             \\ \hline
TL-AMC     & Wang et  al. \cite{wang2020transfer} & The equalized I/Q sequence           & CAE and CNN                        \\ \hline
\end{tabular}
\end{center} 
\end{table}

\subsection{The adopted MIMO system with precoding}
\label{MIMOAMR}
Precoding for MIMO systems has gradually become a research hot spot in recent years with the development of millimeter wave communications technology, and research on the MIMO system incorporating precoding is gaining increased attention. In this paper, we test DL-AMR for the MIMO system containing precoding, and the detailed MIMO system is described as follows.

\begin{figure}[htbp]
	\centering
	\includegraphics[scale=0.5]{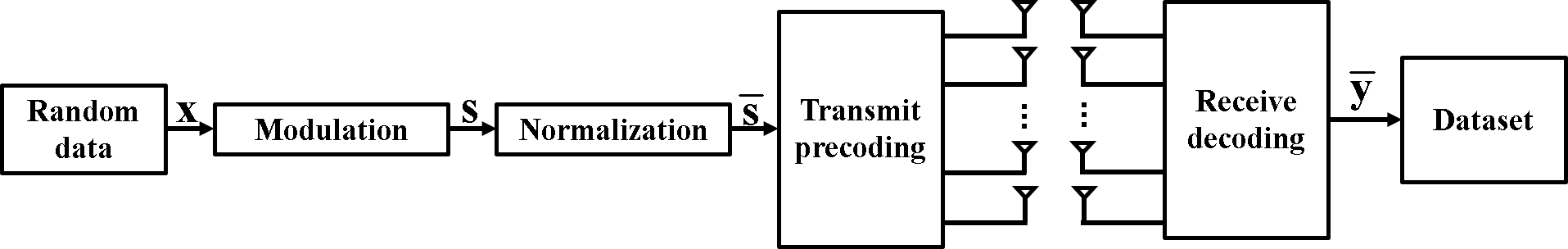}
	\caption{Training data generation for AMR in MIMO system with precoding.}
	\label{FIG:4}
\end{figure}

We consider a typical MIMO system with $Nt$ antennas at the transmitter and $Nr$ antennas at the receiver, as shown in Fig. \ref{FIG:4}. Transmission is over a flat fading channel $H$ ($H \in \mathbb{C}^{Nr \times Nt}$), and both the transmitter and receiver are assumed to have full knowledge of the channel. At the transmitter, the random data ${\textbf{x}} = [{x_1},{x_2},...,{x_k}]$ is modulated into ${\textbf{s}} = [{s_1},{s_2},...,{s_k}]$ using one of the candidate modulation schemes, and then normalized to $\bar{\textbf{s}}$ with unit power, where $k$ is the number of transmitted symbols.

The singular value decomposition (SVD) approach is employed for precoding \cite{SVDprecoding}, and the SVD of $\textbf{H}$ can be written as

\begin{equation}
\label{SVD}
\textbf{H}=\textbf{U}{\Sigma}\textbf{V}^H,
\end{equation}
where $\textbf{U}$ and $\textbf{V}$ are unitary matrices, $\textbf{V}^*$ denotes the transpose conjugate of $\textbf{V}$, and ${\Sigma}$ is a diagonal matrix of the singular values of $\textbf{H}$ sorted in descending order. 

The optimal vectors to be used at the transmitter side and receiver side are the first column of $\textbf{U}$ and $\textbf{V}$ corresponding to the largest singular value of $\textbf{H}$, which are denoted by $\textbf{U}_1$ and $\textbf{V}_1$, respectively. The modulated data $\bar{\textbf{s}}$ is multiplied by the matrix $\textbf{V}_1$ before being transmitted over the channel. In turn, the receiver multiplies the received data by the matrix $\textbf{U}_1^*$. Then, the vector of the received symbol can be expressed as

\begin{equation}
\label{receivesymbol}
\bar{\textbf{y}}=\textbf{U}_1^H(\textbf{H}\textbf{V}_1{\bar{\textbf{s}}}+{\textbf{n}})= {\lambda}_1{\bar{\textbf{s}}}+\bar{\textbf{n}}
\end{equation}
where $\textbf{n}$ is a vector of additive white Gaussian noise (AWGN) at the receiver, $\bar{\textbf{n}} = \textbf{U}^H\textbf{n}$ has the same distribution of the noise $\textbf{n}$, ${\lambda}_1$ is the largest singular value of $\textbf{H}$, and $\bar{\textbf{y}} = [{\bar{y}_1},{\bar{y}_2},...,{\bar{y}_k}]$ indicates the data to be recognized. Due to the extensive works on blind channel estimation \cite{tugnait2001blind,zeng2006blind}, we can assume that the channel state information is perfect at the receiver, thus allowing modulation recognition of the signal after receiving decoding.
\section{Experimental comparison for SISO system}
\label{section:Experimental comparison}

\begin{figure}[htbp]
	\centering
	\includegraphics[scale=0.04]{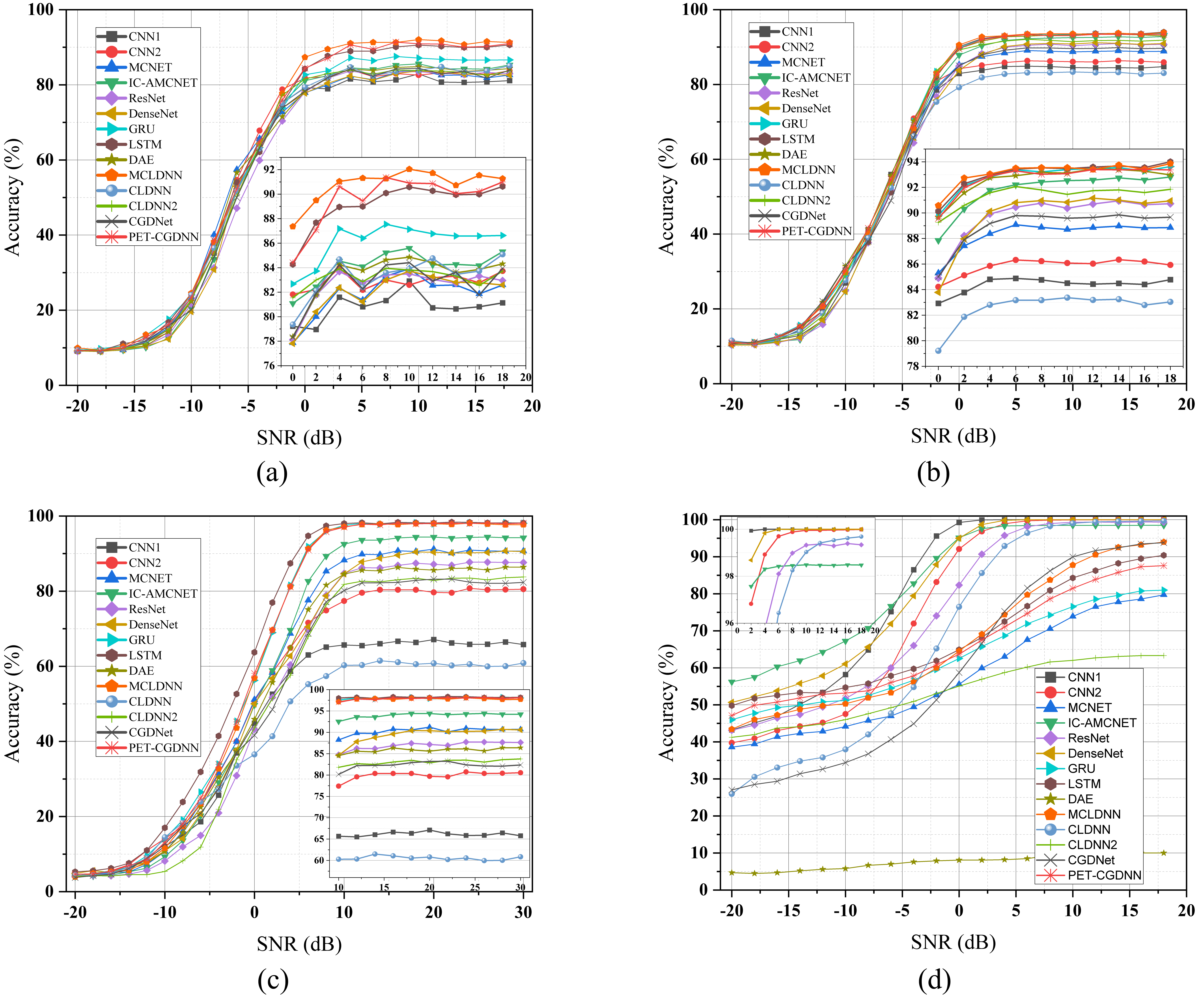}
	\caption{Recognition accuracy comparison of the state-of-the-art models on (a) RML2016.10a, (b) RML2016.10b, (c) RML2018.01a, (d) HisarMod2019.1.}
	\label{FIG:5}
\end{figure}

This section evaluates the state-of-the-art supervised deep learning models for AMR challenges using extensive experiments focusing on model structures, complexity and recognition accuracy, to provide a complete picture of this field. The experiments will further highlight the key unsolved problems of DL-AMR and the promising future research directions.

\subsection{Experimental Setup}
\label{SISOexSet}
We choose the benchmark datasets RML2016.10a, RML2016.10b, RML2018.01a, and HisarMod2019.1 for all experiments. The input data dimension of RML2016.10a and RML2016.10b is 2 × 128, while that of RML2018.01a and HisarMod2019.1 is 2 × 1024. We only adapt the input dimension and output dimension of the models to fit the input data, and keep the same intermediate parameters across all datasets.

We divide each RML dataset into the training set, validation set and test set at the ratio of 6:2:2, while HisarMod2019.1 dataset is divided into the ratio of 8:2:5 due to its data storage format. The categorical cross-entropy is set as the loss function and the Adam algorithm is used as the optimizer. The initial learning rate starts at 0.001 and the batch size is set as 400 in all experiments. If the validation loss does not decrease in 5 epochs, the learning rate will be halved. The training process stops if the validation loss remains stable in 50 epochs. The experiments are implemented using GeForce GTX 1080Ti GPU and Keras with Tensorflow as the backend.

\subsection{Recognition Accuracy}
The experimental accuracy of the 14 models is shown in Fig. \ref{FIG:5}. Under the same experimental parameter settings, the highest recognition accuracy achieved on RML2016.10a, RML2016.10b, RML2018.01a and HisarMod2019.1 datasets are 92.05\% (by MCLDNN at 10dB), 94\% (by LSTM at 18dB), 98.39\% (by LSTM at 22dB), and 100\% (by CNN1 and DenseNet at high SNR), respectively. 

From the overall experimental results on RML datasets using either I/Q or amplitude/phase as input, several AMR models based on RNN, e.g., LSTM, GRU, PET-CGDNN, and MCLDNN, have clear advantages in recognition accuracy. The experimental results show that the temporal features of the modulated signals provide pivotal support to the recognition accuracy on RML datasets, and the spatial features of signals extracted by CNN models are more effective for HisarMod2019.1. Since HisarMod2019.1 is generated by random bit sequences and RML datasets are generated by specific source data (Serial Episode \#1 and Gutenberg works of Shakespeare in ASCII \cite{o2016radio}), RML data samples contain more relevant information on the temporal scale, so the RNN layer, which is more capable of extracting temporal features, can not perform as well on HisarMod2019.1 dataset as it does on RML datasets. On the other hand, the channel conditions for generating RML datasets are more complex (e.g., sample rate offset and center frequency offset), so that the RML data symbols sampled at different time have more relevant features on the temporal scale, and the effective extraction of these features by RNN layers can improve the recognition accuracy.

Note that CNN2 has more convolutional layers than CNN1, and the accuracy on RML datasets suggests that a model with more convolutional layers can achieve the same or better performance on certain tasks with fewer parameters. However, the simplest CNN1 structure achieves better performance on HisarMod2019.1 dataset, implying that the model with more layers or more complex structures (e.g., CLDNN2 and DAE) may fall into local optimality on this relatively more ideal dataset and is prone to degradation problems \cite{he2016deep}. The models based on ResNet and DenseNet, which contain an exceptionally large number of internal parameters and achieve superior performance in computer vision tasks, fail to achieve equivalent performance in RML datasets but perform well on HisarMod2019.1 dataset. Multi-channel inputs, as adopted by MCLDNN, result in higher accuracy on RML datasets than the single-channel structure used by several models as shown in the accuracy figures. 

The validation loss in the training process shows that all the models converge quickly and reach a steady status beyond 40 epochs, among which MCLDNN, PET-CGDNN, LSTM, and GRU achieve the best convergence trends across three RML datasets. Some other models, e.g., DAE, CLDNN, and CLDNN2, have varied performances and significantly different parameters in different datasets, showing their potential improvements in generalisation. Particularly, RML2018.01a dataset is a larger dataset compared with RML2016.10a and RML2016.10b, and a model's hyperparameters could be appropriately increased to benefit from the rich data samples. In order to provide a fair comparison, this paper adopts the same hyperparameters. Interested researchers could explore the effect of hyperparameters based on the trends of experimental results here. 

\subsection{Model complexity and training time}
\begin{table}[htbp]
\caption{Model size and complexity comparison on the four datasets (A: RML2016.10a, B: RML2016.10b, C: RML2018.01a, D: HisarMod2019.1).}
\label{tbl4}
\resizebox{\textwidth}{!}{
\begin{tabular}{c|cccc|cccc|cccc|cccc}
\hline\hline
\textbf{Model} & \multicolumn{4}{c|}{\textbf{\begin{tabular}[c]{@{}c@{}}Learning\\ parameters\end{tabular}}} & \multicolumn{4}{c|}{\textbf{\begin{tabular}[c]{@{}c@{}}Training time\\  (second/epoch)\end{tabular}}} & \multicolumn{4}{c|}{\textbf{\begin{tabular}[c]{@{}c@{}}Training\\ epochs\end{tabular}}} & \multicolumn{4}{c}{\textbf{\begin{tabular}[c]{@{}c@{}}Minimum\\ validation loss\end{tabular}}} \\ \hline
\textbf{Dataset} & \textbf{A} & \textbf{B} & \textbf{C} & \textbf{D}& \textbf{A} & \textbf{B} & \textbf{C} & \textbf{D}& \textbf{A} & \textbf{B} & \textbf{C} & \textbf{D}& \textbf{A} & \textbf{B} & \textbf{C}& \textbf{D}\\ \hline
\textbf{CNN1} & 1592383 & 1592126 & 13064524 & 13065038& \textbf{5} & \textbf{26} & 147 & 173 & 106 & 157 & 166 & 173& 1.1578 & 0.9863 & 1.5525& 0.6128\\ \hline
\textbf{CNN2} & 858123 & 857994 & 1777304 & 1777562& 17 & 98 & 676 & 353 & 132 & 234 & 230 & 228 & 1.1183 & 0.9643 & 1.3608 & 0.7139\\ \hline
\textbf{MCNET} & 121511 & 121226 & 126616& 127386 & 8 & 46 & \textbf{100} & \textbf{43}& 93 & 107 & \textbf{82} &128 & 1.1366 & 0.9590 & 1.2765 & 1.1359\\ \hline
\textbf{IC-AMCNET} & 1264011 & 1263882 & 8605720 & 8605978 & 6 & 34 & 172 & 92 & 262 & 233 & 89 & 250 & 1.1402 & 0.9396 & 1.2725& \textbf{0.5419}\\ \hline
\textbf{ResNet} & 3098283 & 3098154 & 21450040 & 21450298 & 23 & 131 & 1006 & 527 & 133 & 275 & 177 & 247 & 1.1946 & 0.9723 & 1.4008 & 0.8766\\ \hline
\textbf{DenseNet} & 3282603 & 3282474 & 21634360 & 21634618 & 35 & 189 & 1523 & 809 & 312 & 298 & 266 & 192 & 1.1872 & 0.9687 & 1.3697 &0.6184\\ \hline
\textbf{GRU} & 151179 & 151050 & 152856 & 153114 & 9 & 48 & 313 & 182 & 103 & 88 & 162 & 301 & 1.1124 & 0.8901 & \textbf{1.1124} & 0.9689\\ \hline
\textbf{LSTM} & 201099 & 200970 & 202776 & 203034 & 11 & 56 & 497 & 225& \textbf{89} & \textbf{68} & 106 & 115& 1.1004 & 0.8887 & 1.1192 &0.8820\\ \hline
\textbf{DAE} & 1063659 & 1063642&  67125960& 67125994&  7&  39& 301 & 163 & 212 &228 & 139 & 653  & 1.1336 &0.9016  &1.3252  & 1.7937 \\ \hline
\textbf{MCLDNN} & 406199 & 406070 & 407876 & 408134 & 17 & 90 & 662 & 369 & 103 & 69 & 92 & 92 & \textbf{1.0612} & \textbf{0.8851} & 1.1313 & 0.9131\\ \hline
\textbf{CLDNN} & 164433 & 164176 & 884374 &884888 & 8 & 45 & 416 &197 & 148 & 97 & 95 & \textbf{70} & 1.1409 & 1.0210 & 1.9658 & 1.0091\\ \hline
\textbf{CLDNN2} & 517643 & 517514 & 698320 & 698578 & 18 & 97 & 963 & 450& 142 & 191 & 130 & 71& 1.1264 & 0.9278 & 1.4780 & 1.1933\\ \hline
\textbf{CGDNet} & 124933 & 124676 & 665874 & 666388 &  7 & 40 &  237 & 121 & 142&  157&  111& 80 & 1.1473 & 0.9560 & 1.4349 & 1.3479\\ \hline
\textbf{PET-CGDNN} & \textbf{71871} & \textbf{71742} & \textbf{75340} &\textbf{75598} & 6 & 33& 208 & 110 & 97 &  101&  311& 201 & 1.0945 &  0.9002& 1.1185 &0.9107\\ \hline\hline
\end{tabular}}
\end{table}

Five key indicators are selected to measure the complexity and evaluate the training process of the 14 models (Table \ref{tbl4}), including the learning parameters, training time, training epochs, and minimum validation loss. A model's complexity is mainly reflected by the learning parameters, the training speed is indicated by the training time and training epoch, while the minimum validation loss demonstrates the convergence trend. 

Due to the local connection and weight-sharing characteristics of CNN, it is more computationally efficient. Although some CNN models have more learnable parameters, such as CNN1 and IC-AMCNET, their training time in one epoch is shorter. It is worth noting that MCLDNN has the minimum validation loss on RML2016.10a and RML2016.10b datasets, which contributes to the high recognition accuracy. PET-CGDNN has the least learning parameters and thus requires less memory than others. CNN-based models generally have 
lower validation loss on HisarMod2019.1 dataset, indicating that CNN models are better suited to this simpler dataset. The high validation loss of MCNET on HisarMod2019.1 dataset suggests that using fewer parameters and a more complex structure on this dataset with a larger input dimension tends to cause poorer model convergence. 

\subsection{Confusion matrices analysis}
\begin{figure}[htbp]
	\centering
	\includegraphics[scale=0.14]{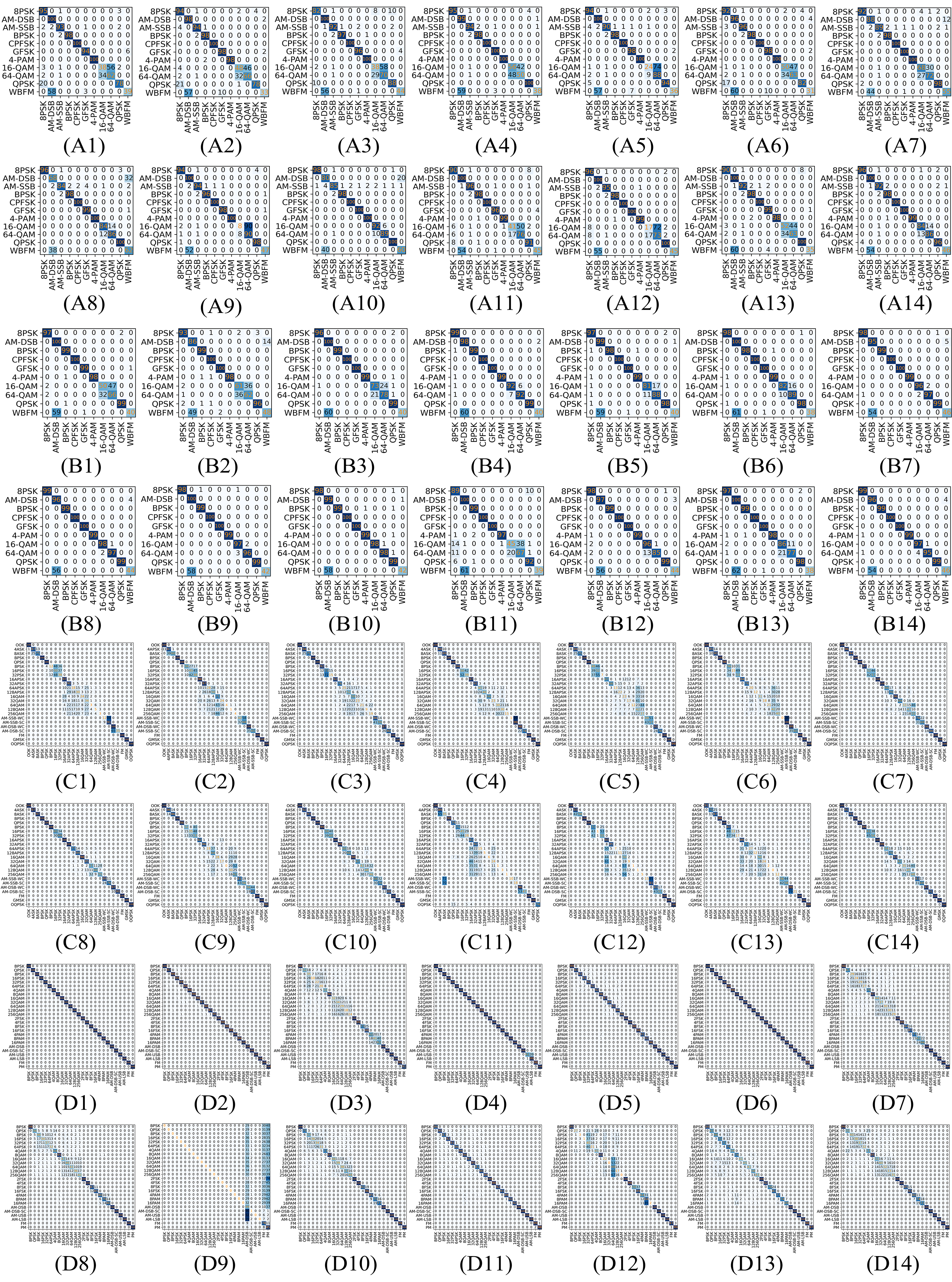}
	\caption{Confusion matrices. A, B and C represent the confusion matrices obtained on the RML2016.10a, RML2016.10b, and RML2018.01a, respectively. The numerical indexes 1 - 14 denote CNN1, CNN2, MCNET, IC-AMCNET, ResNet, DenseNet, GRU, LSTM, DAE, MCLDNN, CLDNN, CLDNN2, CGDNet, PET-CGDNN.}
	\label{FIG:6}
\end{figure}

The confusion matrices of the 14 models on the three datasets provide important insights on the main modulation types that cause significant classification errors. Fig. \ref{FIG:6} selects the typical confusion matrices when SNR is moderate, e.g., at 4dB, across 14 models. The vertical axis on each matrix denotes the true labels while the horizontal axis denotes the predicted labels. We summarize the key observations below: 
On RML2016.10a (Fig. \ref{FIG:6}: (A1) - (A14)): 
\begin{itemize}
    \item  All models confuse a limited number of samples modulated with 16QAM and 64QAM, caused by the overlapping constellation points between these two modulation schemes. MCLDNN performs best owing partially to the multi-channel structure that differentiates the constellation points that are not completely overlapped.
    
    \item A significant number of WBFM modulation data could be classified as AM-DSB by all models. The data of WBFM and AM-DSB both come from a speech signal with some interludes and off-time \cite{o2016convolutional}, leading to classification errors.
\end{itemize}

On RML2016.10b (Fig. \ref{FIG:6}: (B1) - (B14)): 
\begin{itemize}
    \item  Similar results were obtained as RML2016.10a because the two datasets have similar data inputs as described before. For example, the confusion between WBFM and AM-DSB still exists.
    
    \item The 16QAM and 64QAM confusion problem on RML2016.10a is slightly alleviated in most models due to the increased number of samples provided by RML2016.10b dataset. 
\end{itemize}
On RML2018.01a (Fig. \ref{FIG:6}: (C1) - (C14)): 
\begin{itemize}
    \item Because more modulation schemes are introduced, more confusion types are observed in almost all the 14 models, e.g., confusion between 4ASK and 8ASK, 16PSK and 32PSK, 64APSK and 128APSK, QAM schemes (32QAM, 64QAM, 128QAM, and 256QAM), AM-DSB-WC and AM-DSB-SC.
    
    \item Data with similar modulation types are more likely to cause misjudgment if they are under strong noise and interference. The increase of SNR could contribute to decreased confusion and improved recognition accuracy. 
\end{itemize}
On HisarMod2019.1 (Fig. \ref{FIG:6}: (D1) - (D14)): 
\begin{itemize}
    \item The confusion problem is not as severe as the results on the RML datasets, e.g., CNN1 has no recognition error, and the models which have confusion problems mainly exist in the same class of modulation schemes, e.g., PSK, QAM, and PAM.
    
    \item DAE model has serious confusion problems on this dataset, with the output modulation schemes distributed in AM-DSB, AM-DSB-SC, FM, and PM, but still maintains some ability to identify FM and PM modulation schemes.
\end{itemize}

\section{Experimental comparison for the adopted MIMO system}
Our main objective is to verify the feasibility of applying DL-AMR in the MIMO system with precoding, rather than to propose a new deep learning model for modulation recognition. Therefore, four representative models are compared in the experiments in this section, including a CNN model \cite{wang2020deep}, an RNN-based LSTM model \cite{rajendran2018deep}, an RNN-based GRU model \cite{hong2017automatic}, and a hybrid CLDNN model \cite{liu2017deep}. This study aims to provide some baseline results for future research to make further improvements and innovations based on our experimental results and open source code.

The dataset is generated by MATLAB and then Keras with Tensorflow as the backend is used to train the modulation recognition model. The data are modulated using different modulation methods, including `2PSK`, `QPSK`, `8PSK`, `16QAM`, `64QAM`, and `128QAM`. The number of transmitted symbols per sample is 128 ($k$=128), and we prepare 500 samples per SNR per scheme, which are divided into three parts for training, validation, and testing by the ratio of 6:2:2. The batchsize is set to 128, and the other parameters are set as the same as the SISO model experiments in Section \ref{SISOexSet}.

\begin{figure}[htbp]
	\centering
	\includegraphics[scale=.034]{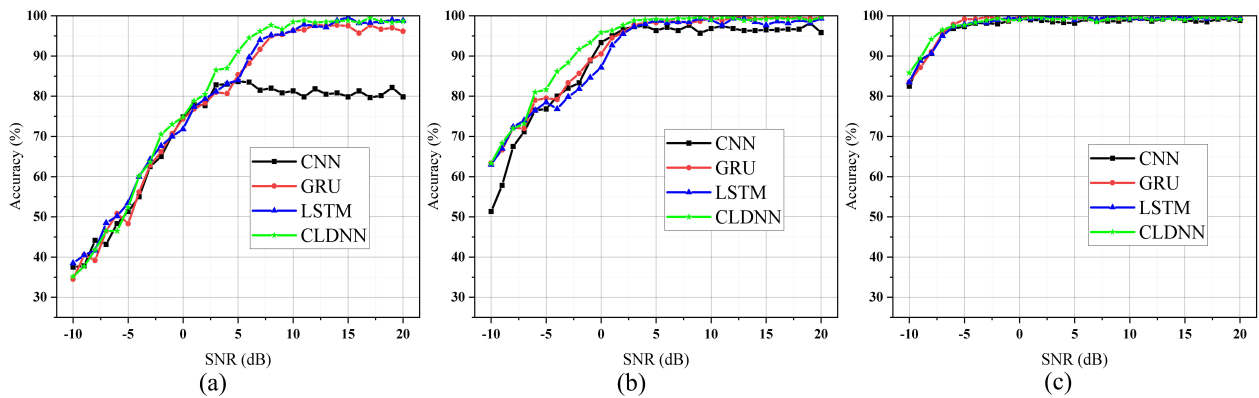}
	\caption{Recognition accuracy comparison of four DL models on the MIMO system containing precoding. (a) $Nt$ = 4, $Nr$ = 2, (b) $Nt$ = 16, $Nr$ = 4, (c) $Nt$ = 64, $Nr$ = 16.}
	\label{FIG:7}
\end{figure}

We conducted experiments with three different antenna configurations: 4*2 ($Nt$ = 4,$Nr$ = 2), 16*4 ($Nt$ = 16,$Nr$ = 4), and 64*16 ($Nt$ = 64,$Nr$ = 16). The performance of different deep learning models with different antenna configurations is shown in Fig. \ref{FIG:7}.

From Fig. \ref{FIG:7}, the CLDNN model has a higher overall recognition accuracy and the CNN model has a lower overall recognition accuracy. The gain made by the increased number of antennas leads to the improved recognition accuracy of each model. The experimental results exhibit that DL-AMR is feasible for the MIMO system with precoding, and SISO's DL-AMR method can be applied to such MIMO systems, since their input data contain only one complex data stream which requires similar network structures. The impact of other factors, such as channel estimation error, precoding schemes, etc, is worthy to be further investigated. 

The AMR of MIMO systems differs somewhat from that of SISO systems, and the main difference is the number of data streams to be processed. In MIMO systems, one way is to perform modulation recognition directly on the data received from multiple antennas, and the data samples have an extra antenna dimension of $Nr$. The data samples received from multiple antennas can be recognized by cooperative recognition methods \cite{wang2020deep} or 3D convolution-based approach \cite{huynh2022mimo} for modulation recognition. Alternatively, the received data from multiple antennas can be processed at the receiver, e.g., by ZF equalization \cite{wang2020automatic} or precoding, to convert the received data from multiple antennas to a single data stream, ensuring the similar DL-AMR method applied for SISO systems can be applied for MIMO systems.

\section{Challenges and future directions}
\label{section:Challenges and future directions}
As described in the previous sections, DL shows great potential for AMR. However, there are still several unsolved problems that are worthy of further study. This section discusses the important problems and opportunities in this area from three perspectives, focusing on model innovation, data and datasets, and model optimization and visualisation.

\subsection{Designing novel DL-AMR models}

\emph{Novel deep learning models specialised for modulation inference}: According to the existing research, CNN-based spatial feature extraction and RNN-based temporal feature extraction have led to the breakthroughs on AMR. Other types of neural network structures such as generative adversarial networks (GAN) \cite{2018Digital}, attention mechanism \cite{chen2020novel,ShishengHu,9467342,9672167,9682126}, and transformer, which have been shown to achieve good performance in specific fields, could be further exploited for AMR. For example, GAN can be used as a generator to expand the training set which can be regarded as a data augmentation method. The attention mechanism can allow the model to focus on relevant features and accelerate the training process. Considering the sequential characteristics of AMR signals, the transformer models, introduced in natural language processing, could further enhance the AMR performance. Joint signal processing and model design would be another promising future direction \cite{yashashwi2018learnable,zhang2021efficient}. 

\emph{Model compression and simplification}: For AMR challenges, and even the wider deep learning community, achieving higher accuracy is often based on stacking more layers and deepening the model. While such strategies may lead to feasible models in many offline tasks, AMR often requires online processing and will encounter excessive delays if a model is too complex. The high computational complexity also prevents their application in resource-constrained devices such as many internet-of-thing (IoT) devices that have limited memory, computing power, and energy, while efficient DL-AMR models with low complexity and lightweight property can meet the requirement of next generation mobile communication systems, such as B5G systems, which have greater demand for extremely reliable and low-latency \cite{9525144}. In the machine learning community, model compression and simplification methods such as pruning \cite{lin2020improved,wang2020lightamc,accesspruning}, knowledge distillation \cite{2020Cross}, and weight quantization have been shown to successfully decrease model complexity significantly while maintaining the same level of performance, hence could be exploited in DL-AMR.

\emph{Model generalisation on different channel conditions}: Most AMR deep learning models are trained with the data generated under specific channel conditions. To apply such models in practical systems would require models to have exceptional generality and robustness. To improve the generality of a model for different channel conditions at the designing stage is a topic worthy of study in the future. One potential method is transfer learning \cite{wang2020transfer}. For example, a DL-AMR model can be trained on offline datasets such as RML2016.10a, and fine-tuned using online data obtained in real-world by adopting the transfer learning strategy. If no training data is available or only a few training samples are available for some signal classes, then zero-shot \cite{9392373} and few-shot \cite{9650842,9245503,9676419} learning are possible solutions for signal recognition. To improve the robustness of a model, methods that can withstand adversarial attacks need to be further investigated \cite{9542973}.

\subsection{Benchmark AMR models, datasets, and data challenges}

\emph{Open-source datasets and benchmark models}: There are only a few open-source datasets available for AMR so far, i.e., RadioML2016.10a, RadioML2016.10b, RadioML2018.01a and HisarMod2019.1. Many research generate their own datasets and train their own models using such datasets, lacking solid common benchmarks for performance comparison and further improvement. More high-quality AMR datasets (similar to ImageNet in  computer vision) and a unified benchmark paradigm will be a challenging but rewarding direction, while data cleaning and augmentation may also be one major challenge \cite{DataAugmentation}. Furthermore, with the wide deployment of MIMO systems, it is also necessary to get the corresponding standard datasets which can be used in AMR research for MIMO systems \cite{9405666}.

\emph{Multi-channel inputs and parallel processing}: Pioneering work on multi-channel inputs and feature fusion, e.g., MCLDNN \cite{xu2020spatiotemporal}, CNN-LSTM \cite{zhang2020automatic}, and WSMF \cite{AMCmulti} has achieved higher accuracy in AMR than the other sophisticated models with single inputs. The rich and often complementary information provided by multiple channels could enable models to learn key distinguishable features, but it also introduces increased computational complexity and potentially poor convergence. The parallel nature of such a method could benefit from high-performance computing architectures based on parallel processing such as graphics processing units (GPUs), which can potentially reduce the computational complexity by many folds. To address the convergence challenge, new loss functions or weight adjustments based on different data types can be exploited. 

\emph{Semi-supervised methods}: Unsupervised learning methods such as DBN \cite{mendis2019deep} and autoencoder \cite{ali2017k} have been exploited in AMR, but have not achieved comparatively distinctive results. Most successful state-of-the-art models are supervised models, which require a large amount of labeled data. Building a high-quality labeled dataset often needs significant investment in resources and time. Semi-supervised models \cite{9383106,wang2020transfer}, which are trained on partially labeled samples or a small number of samples, have achieved breakthroughs in a wide range of applications such as classification and detection, and could therefore be applied in AMR to mitigate the needs for labeled data from different channel conditions.

\subsection{Model optimization and visualisation}
\emph{Fusion of deep learning models with expert knowledge}: Deep learning models learn representative features purely from data, while expert knowledge of wireless communications systems has proved to be highly reliable. The fusion of deep learning models with expert knowledge (e.g., high order cumulants (HOC), handcrafted features, spectrum, constellation diagrams, etc.), could lead to a potential breakthrough in AMR \cite{BJD}. Furthermore, signal processing techniques such as filtering \cite{9605585,zeng2019spectrum} and denoising \cite{9440746} can also help retain distinctive data features and improve system performance.

\emph{AMR model visualisation}: The low SNR of the input signal in AMR makes the features extracted by a DL model difficult to assess visually, which affects the design and development of new models. On the one hand, good visualization techniques help design models and optimize parameters thus achieving better performance \cite{huang2020visualizing}. On the other hand, visualization can provide the explainability of DL-AMR models, from which we can know what kinds of signal features should be extracted and chosen for classification. 
\section{Conclusion}
\label{section:Conclusion}
This paper reviews the emerging interdisciplinary topic of DL-AMR, which also analyzes the characteristics of benchmark models and discusses the available datasets. We have further identified the unresolved problems and proposed possible solutions through extensive experimental comparisons, which have provided detailed insights of this topic for future research. With the ubiquitous deployment of cognitive communications systems and distributed networked devices in future B5G/6G networks, DL-AMR will have an increasingly important role to play.

\section*{Declaration of competing interest}
The authors declare that they have no known competing financial interests or personal relationships that could have appeared to influence the work reported in this paper.

\section*{Acknowledgments}
This work was supported by the National Natural Science Foundation of China under grant 61871096, and in part by the National Key R\&D Program of China under grant 2018YFB2101300.

\bibliographystyle{cas-model2-names}
\bibliography{main}

\begin{thebibliography}{94}
\expandafter\ifx\csname natexlab\endcsname\relax\def\natexlab#1{#1}\fi
\providecommand{\url}[1]{\texttt{#1}}
\providecommand{\href}[2]{#2}
\providecommand{\path}[1]{#1}
\providecommand{\DOIprefix}{doi:}
\providecommand{\ArXivprefix}{arXiv:}
\providecommand{\URLprefix}{URL: }
\providecommand{\Pubmedprefix}{pmid:}
\providecommand{\doi}[1]{\href{http://dx.doi.org/#1}{\path{#1}}}
\providecommand{\Pubmed}[1]{\href{pmid:#1}{\path{#1}}}
\providecommand{\bibinfo}[2]{#2}
\ifx\xfnm\relax \def\xfnm[#1]{\unskip,\space#1}\fi
\bibitem[{Ali and Fan(2017)}]{ali2017k}
\bibinfo{author}{Ali, A.}, \bibinfo{author}{Fan, Y.}, \bibinfo{year}{2017}.
\newblock \bibinfo{title}{{$k$-S}parse autoencoder-based automatic modulation
  classification with low complexity}.
\newblock \bibinfo{journal}{IEEE Commun. Lett.} \bibinfo{volume}{21},
  \bibinfo{pages}{2162--2165}.
\bibitem[{Ali and Yangyu(2017)}]{SpaseAli}
\bibinfo{author}{Ali, A.}, \bibinfo{author}{Yangyu, F.}, \bibinfo{year}{2017}.
\newblock \bibinfo{title}{Automatic modulation classification using deep
  learning based on sparse autoencoders with nonnegativity constraints}.
\newblock \bibinfo{journal}{IEEE Signal Process. Lett.} \bibinfo{volume}{24},
  \bibinfo{pages}{1626--1630}.
\bibitem[{Bhatti et~al.(2021)Bhatti, Khan, Selim and Paisana}]{9395503}
\bibinfo{author}{Bhatti, F.A.}, \bibinfo{author}{Khan, M.J.},
  \bibinfo{author}{Selim, A.}, \bibinfo{author}{Paisana, F.},
  \bibinfo{year}{2021}.
\newblock \bibinfo{title}{Shared spectrum monitoring using deep learning}.
\newblock \bibinfo{journal}{IEEE Trans. Cogn. Commun. Netw.}
  \bibinfo{volume}{7}, \bibinfo{pages}{1171--1185}.
\bibitem[{Chan et~al.(1985)Chan, Gadbois and Yansouni}]{chan1985identification}
\bibinfo{author}{Chan, Y.}, \bibinfo{author}{Gadbois, L.},
  \bibinfo{author}{Yansouni, P.}, \bibinfo{year}{1985}.
\newblock \bibinfo{title}{Identification of the modulation type of a signal},
  in: \bibinfo{booktitle}{Proc. IEEE Int. Conf. Acoust., Speech, Signal
  Process.}, pp. \bibinfo{pages}{838--841}.
\bibitem[{Chang et~al.(2022)Chang, Huang, Zhang, Feng and Liu}]{9462447}
\bibinfo{author}{Chang, S.}, \bibinfo{author}{Huang, S.},
  \bibinfo{author}{Zhang, R.}, \bibinfo{author}{Feng, Z.},
  \bibinfo{author}{Liu, L.}, \bibinfo{year}{2022}.
\newblock \bibinfo{title}{Multitask-learning-based deep neural network for
  automatic modulation classification}.
\newblock \bibinfo{journal}{IEEE Internet Things J.} \bibinfo{volume}{9},
  \bibinfo{pages}{2192--2206}.
\bibitem[{Chen et~al.(2020)Chen, Zhang, He, Nie and Zhang}]{chen2020novel}
\bibinfo{author}{Chen, S.}, \bibinfo{author}{Zhang, Y.}, \bibinfo{author}{He,
  Z.}, \bibinfo{author}{Nie, J.}, \bibinfo{author}{Zhang, W.},
  \bibinfo{year}{2020}.
\newblock \bibinfo{title}{A novel attention cooperative framework for automatic
  modulation recognition}.
\newblock \bibinfo{journal}{IEEE Access} \bibinfo{volume}{8},
  \bibinfo{pages}{15673--15686}.
\bibitem[{Chen et~al.(2021)Chen, Cui, Xiang, Qiu, Huang, Zheng, Chen, Xuan and
  Yang}]{9580446}
\bibinfo{author}{Chen, Z.}, \bibinfo{author}{Cui, H.}, \bibinfo{author}{Xiang,
  J.}, \bibinfo{author}{Qiu, K.}, \bibinfo{author}{Huang, L.},
  \bibinfo{author}{Zheng, S.}, \bibinfo{author}{Chen, S.},
  \bibinfo{author}{Xuan, Q.}, \bibinfo{author}{Yang, X.}, \bibinfo{year}{2021}.
\newblock \bibinfo{title}{{SigNet}: A novel deep learning framework for radio
  signal classification}.
\newblock \bibinfo{journal}{IEEE Trans. Cogn. Commun. Netw.} ,
  \bibinfo{pages}{1--1}.
\bibitem[{Dai et~al.(2016)Dai, Zhang and Hong}]{AF-SAE}
\bibinfo{author}{Dai, A.}, \bibinfo{author}{Zhang, H.}, \bibinfo{author}{Hong,
  S.}, \bibinfo{year}{2016}.
\newblock \bibinfo{title}{Automatic modulation classification using stacked
  sparse auto-encoders}, in: \bibinfo{booktitle}{Proc. IEEE 13th Int. Conf.
  Signal Process.}
\bibitem[{Dobre et~al.(2007)Dobre, Abdi, Bar-Ness and Su}]{dobre2007survey}
\bibinfo{author}{Dobre, O.A.}, \bibinfo{author}{Abdi, A.},
  \bibinfo{author}{Bar-Ness, Y.}, \bibinfo{author}{Su, W.},
  \bibinfo{year}{2007}.
\newblock \bibinfo{title}{Survey of automatic modulation classification
  techniques: classical approaches and new trends}.
\newblock \bibinfo{journal}{IET commun.} \bibinfo{volume}{1},
  \bibinfo{pages}{137--156}.
\bibitem[{Dong et~al.(2021a)Dong, Jiang, Cheng and Shi}]{9383106}
\bibinfo{author}{Dong, Y.}, \bibinfo{author}{Jiang, X.},
  \bibinfo{author}{Cheng, L.}, \bibinfo{author}{Shi, Q.},
  \bibinfo{year}{2021}a.
\newblock \bibinfo{title}{{SSRCNN}: A semi-supervised learning framework for
  signal recognition}.
\newblock \bibinfo{journal}{IEEE Trans. Cogn. Commun. Netw.}
  \bibinfo{volume}{7}, \bibinfo{pages}{780--789}.
\bibitem[{Dong et~al.(2021b)Dong, Jiang, Zhou, Lin and Shi}]{9392373}
\bibinfo{author}{Dong, Y.}, \bibinfo{author}{Jiang, X.}, \bibinfo{author}{Zhou,
  H.}, \bibinfo{author}{Lin, Y.}, \bibinfo{author}{Shi, Q.},
  \bibinfo{year}{2021}b.
\newblock \bibinfo{title}{{SR2CNN}: Zero-shot learning for signal recognition}.
\newblock \bibinfo{journal}{IEEE Trans. Signal Process.} \bibinfo{volume}{69},
  \bibinfo{pages}{2316--2329}.
\bibitem[{Dulek(2017)}]{dulek2017online}
\bibinfo{author}{Dulek, B.}, \bibinfo{year}{2017}.
\newblock \bibinfo{title}{Online hybrid likelihood based modulation
  classification using multiple sensors}.
\newblock \bibinfo{journal}{IEEE Trans. Wireless Commun.} \bibinfo{volume}{16},
  \bibinfo{pages}{4984--5000}.
\bibitem[{Ghasemzadeh et~al.(2020)Ghasemzadeh, Banerjee, Hempel and
  Sharif}]{9188007}
\bibinfo{author}{Ghasemzadeh, P.}, \bibinfo{author}{Banerjee, S.},
  \bibinfo{author}{Hempel, M.}, \bibinfo{author}{Sharif, H.},
  \bibinfo{year}{2020}.
\newblock \bibinfo{title}{A novel deep learning and polar transformation
  framework for an adaptive automatic modulation classification}.
\newblock \bibinfo{journal}{IEEE Trans. Veh. Technol.} \bibinfo{volume}{69},
  \bibinfo{pages}{13243--13258}.
\bibitem[{Ghasemzadeh et~al.(2021)Ghasemzadeh, Hempel, Banerjee and
  Sharif}]{9405666}
\bibinfo{author}{Ghasemzadeh, P.}, \bibinfo{author}{Hempel, M.},
  \bibinfo{author}{Banerjee, S.}, \bibinfo{author}{Sharif, H.},
  \bibinfo{year}{2021}.
\newblock \bibinfo{title}{A spatial-diversity {MIMO} dataset for {RF} signal
  processing research}.
\newblock \bibinfo{journal}{IEEE Trans. Instrum. Meas.} \bibinfo{volume}{70},
  \bibinfo{pages}{1--10}.
\bibitem[{Ghasemzadeh et~al.(2022)Ghasemzadeh, Hempel and Sharif}]{9672088}
\bibinfo{author}{Ghasemzadeh, P.}, \bibinfo{author}{Hempel, M.},
  \bibinfo{author}{Sharif, H.}, \bibinfo{year}{2022}.
\newblock \bibinfo{title}{{GS-QRNN}: A high-efficiency automatic modulation
  classifier for cognitive radio {IoT}}.
\newblock \bibinfo{journal}{IEEE Internet Things J.} , \bibinfo{pages}{1--1}.
\bibitem[{Hanna et~al.(2022)Hanna, Dick and Cabric}]{9605585}
\bibinfo{author}{Hanna, S.}, \bibinfo{author}{Dick, C.},
  \bibinfo{author}{Cabric, D.}, \bibinfo{year}{2022}.
\newblock \bibinfo{title}{Signal processing-based deep learning for blind
  symbol decoding and modulation classification}.
\newblock \bibinfo{journal}{IEEE J. Sel. Areas Commun.} \bibinfo{volume}{40},
  \bibinfo{pages}{82--96}.
\bibitem[{Hazza et~al.(2013)Hazza, Shoaib, Alshebeili and
  Fahad}]{hazza2013overview}
\bibinfo{author}{Hazza, A.}, \bibinfo{author}{Shoaib, M.},
  \bibinfo{author}{Alshebeili, S.A.}, \bibinfo{author}{Fahad, A.},
  \bibinfo{year}{2013}.
\newblock \bibinfo{title}{An overview of feature-based methods for digital
  modulation classification}, in: \bibinfo{booktitle}{Proc. 1st Int. Conf.
  Commun., Signal Process., Appl.}, pp. \bibinfo{pages}{1--6}.
\bibitem[{He et~al.(2016)He, Zhang, Ren and Sun}]{he2016deep}
\bibinfo{author}{He, K.}, \bibinfo{author}{Zhang, X.}, \bibinfo{author}{Ren,
  S.}, \bibinfo{author}{Sun, J.}, \bibinfo{year}{2016}.
\newblock \bibinfo{title}{Deep residual learning for image recognition}, in:
  \bibinfo{booktitle}{Proc. IEEE Conf. Comput. Vis. Pattern Recog.}, pp.
  \bibinfo{pages}{770--778}.
\bibitem[{Hermawan et~al.(2020)Hermawan, Ginanjar, Kim and
  Lee}]{hermawan2020cnn}
\bibinfo{author}{Hermawan, A.P.}, \bibinfo{author}{Ginanjar, R.R.},
  \bibinfo{author}{Kim, D.S.}, \bibinfo{author}{Lee, J.M.},
  \bibinfo{year}{2020}.
\newblock \bibinfo{title}{{CNN}-based automatic modulation classification for
  beyond 5{G} communications}.
\newblock \bibinfo{journal}{IEEE Commun. Lett.} \bibinfo{volume}{24},
  \bibinfo{pages}{1038--1041}.
\bibitem[{Hiremath et~al.(2019)Hiremath, Behura, Kedia, Deshmukh and
  Patra}]{hiremath2019deep}
\bibinfo{author}{Hiremath, S.M.}, \bibinfo{author}{Behura, S.},
  \bibinfo{author}{Kedia, S.}, \bibinfo{author}{Deshmukh, S.},
  \bibinfo{author}{Patra, S.K.}, \bibinfo{year}{2019}.
\newblock \bibinfo{title}{Deep learning-based modulation classification using
  time and stockwell domain channeling}, in: \bibinfo{booktitle}{Proc. Nat.
  Conf. Commun.}, pp. \bibinfo{pages}{1--6}.
\bibitem[{Hong et~al.(2017)Hong, Zhang and Xu}]{hong2017automatic}
\bibinfo{author}{Hong, D.}, \bibinfo{author}{Zhang, Z.}, \bibinfo{author}{Xu,
  X.}, \bibinfo{year}{2017}.
\newblock \bibinfo{title}{Automatic modulation classification using recurrent
  neural networks}, in: \bibinfo{booktitle}{Proc. IEEE Int. Conf. Comput.
  Commun.}, pp. \bibinfo{pages}{695--700}.
\bibitem[{Hong and Ho(1999)}]{hong1999identification}
\bibinfo{author}{Hong, L.}, \bibinfo{author}{Ho, K.}, \bibinfo{year}{1999}.
\newblock \bibinfo{title}{Identification of digital modulation types using the
  wavelet transform}, in: \bibinfo{booktitle}{Proc. IEEE Military Commun.
  Conf.}, pp. \bibinfo{pages}{427--431}.
\bibitem[{Hu et~al.(2019)Hu, Pei, Liang and Liang}]{ShishengHu}
\bibinfo{author}{Hu, S.}, \bibinfo{author}{Pei, Y.}, \bibinfo{author}{Liang,
  P.P.}, \bibinfo{author}{Liang, Y.C.}, \bibinfo{year}{2019}.
\newblock \bibinfo{title}{Deep neural network for robust modulation
  classification under uncertain noise conditions}.
\newblock \bibinfo{journal}{IEEE Trans. Veh. Technol.} \bibinfo{volume}{PP},
  \bibinfo{pages}{1--1}.
\bibitem[{Huang et~al.(2020a)Huang, Pan, Zhang, Qian, Gao and
  Wu}]{DataAugmentation}
\bibinfo{author}{Huang, L.}, \bibinfo{author}{Pan, W.}, \bibinfo{author}{Zhang,
  Y.}, \bibinfo{author}{Qian, L.}, \bibinfo{author}{Gao, N.},
  \bibinfo{author}{Wu, Y.}, \bibinfo{year}{2020}a.
\newblock \bibinfo{title}{Data augmentation for deep learning-based radio
  modulation classification}.
\newblock \bibinfo{journal}{IEEE Access} \bibinfo{volume}{8},
  \bibinfo{pages}{1498--1506}.
\bibitem[{Huang et~al.(2020b)Huang, Zhang, Pan, Chen, Qian and
  Wu}]{huang2020visualizing}
\bibinfo{author}{Huang, L.}, \bibinfo{author}{Zhang, Y.}, \bibinfo{author}{Pan,
  W.}, \bibinfo{author}{Chen, J.}, \bibinfo{author}{Qian, L.P.},
  \bibinfo{author}{Wu, Y.}, \bibinfo{year}{2020}b.
\newblock \bibinfo{title}{Visualizing deep learning-based radio modulation
  classifier}.
\newblock \bibinfo{journal}{IEEE Trans. Cogn. Commun. Netw.}
  \bibinfo{volume}{7}, \bibinfo{pages}{47--58}.
\bibitem[{Huang et~al.(2019a)Huang, Chai, Li, Zhang and Feng}]{CCNN}
\bibinfo{author}{Huang, S.}, \bibinfo{author}{Chai, L.}, \bibinfo{author}{Li,
  Z.}, \bibinfo{author}{Zhang, D.}, \bibinfo{author}{Feng, Z.},
  \bibinfo{year}{2019}a.
\newblock \bibinfo{title}{Automatic modulation classification using compressive
  convolutional neural network}.
\newblock \bibinfo{journal}{IEEE Access} \bibinfo{volume}{PP},
  \bibinfo{pages}{1--1}.
\bibitem[{Huang et~al.(2019b)Huang, Jiang, Gao, Feng and Zhang}]{CFCN}
\bibinfo{author}{Huang, S.}, \bibinfo{author}{Jiang, Y.}, \bibinfo{author}{Gao,
  Y.}, \bibinfo{author}{Feng, Z.}, \bibinfo{author}{Zhang, P.},
  \bibinfo{year}{2019}b.
\newblock \bibinfo{title}{Automatic modulation classification using contrastive
  fully convolutional network}.
\newblock \bibinfo{journal}{IEEE Wireless Commun. Lett.} ,
  \bibinfo{pages}{1--1}.
\bibitem[{Huynh-The et~al.(2020)Huynh-The, Hua, Pham and Kim}]{huynh2020mcnet}
\bibinfo{author}{Huynh-The, T.}, \bibinfo{author}{Hua, C.H.},
  \bibinfo{author}{Pham, Q.V.}, \bibinfo{author}{Kim, D.S.},
  \bibinfo{year}{2020}.
\newblock \bibinfo{title}{{MCN}et: {A}n efficient {CNN} architecture for robust
  automatic modulation classification}.
\newblock \bibinfo{journal}{IEEE Commun. Lett.} \bibinfo{volume}{24},
  \bibinfo{pages}{811--815}.
\bibitem[{Huynh-The et~al.(2022)Huynh-The, Nguyen, Pham, Kim and
  Da~Costa}]{huynh2022mimo}
\bibinfo{author}{Huynh-The, T.}, \bibinfo{author}{Nguyen, T.V.},
  \bibinfo{author}{Pham, Q.V.}, \bibinfo{author}{Kim, D.S.},
  \bibinfo{author}{Da~Costa, D.B.}, \bibinfo{year}{2022}.
\newblock \bibinfo{title}{{MIMO-OFDM} modulation classification using
  three-dimensional convolutional network}.
\newblock \bibinfo{journal}{IEEE Trans. Veh. Technol.} .
\bibitem[{Huynh-The et~al.(2021)Huynh-The, Pham, Nguyen, Nguyen, Ruby, Zeng and
  Kim}]{9576081}
\bibinfo{author}{Huynh-The, T.}, \bibinfo{author}{Pham, Q.V.},
  \bibinfo{author}{Nguyen, T.V.}, \bibinfo{author}{Nguyen, T.T.},
  \bibinfo{author}{Ruby, R.}, \bibinfo{author}{Zeng, M.}, \bibinfo{author}{Kim,
  D.S.}, \bibinfo{year}{2021}.
\newblock \bibinfo{title}{Automatic modulation classification: A deep
  architecture survey}.
\newblock \bibinfo{journal}{IEEE Access} \bibinfo{volume}{9},
  \bibinfo{pages}{142950--142971}.
\bibitem[{Jdid et~al.(2021)Jdid, Hassan, Dayoub, Lim and Mokayef}]{9399081}
\bibinfo{author}{Jdid, B.}, \bibinfo{author}{Hassan, K.},
  \bibinfo{author}{Dayoub, I.}, \bibinfo{author}{Lim, W.H.},
  \bibinfo{author}{Mokayef, M.}, \bibinfo{year}{2021}.
\newblock \bibinfo{title}{Machine learning based automatic modulation
  recognition for wireless communications: A comprehensive survey}.
\newblock \bibinfo{journal}{IEEE Access} \bibinfo{volume}{9},
  \bibinfo{pages}{57851--57873}.
\bibitem[{Ke and Vikalo(2022)}]{9487492}
\bibinfo{author}{Ke, Z.}, \bibinfo{author}{Vikalo, H.}, \bibinfo{year}{2022}.
\newblock \bibinfo{title}{Real-time radio technology and modulation
  classification via an {LSTM} auto-encoder}.
\newblock \bibinfo{journal}{IEEE Trans. Wireless Commun.} \bibinfo{volume}{21},
  \bibinfo{pages}{370--382}.
\bibitem[{Kumar et~al.(2020)Kumar, Sheoran, Jajoo and Yadav}]{CNN3}
\bibinfo{author}{Kumar, Y.}, \bibinfo{author}{Sheoran, M.},
  \bibinfo{author}{Jajoo, G.}, \bibinfo{author}{Yadav, S.K.},
  \bibinfo{year}{2020}.
\newblock \bibinfo{title}{Automatic modulation classification based on
  constellation density using deep learning}.
\newblock \bibinfo{journal}{IEEE Commun. Lett.} \bibinfo{volume}{PP},
  \bibinfo{pages}{1--1}.
\bibitem[{Lebrun et~al.(2005)Lebrun, Gao and Faulkner}]{SVDprecoding}
\bibinfo{author}{Lebrun, G.}, \bibinfo{author}{Gao, J.},
  \bibinfo{author}{Faulkner, M.}, \bibinfo{year}{2005}.
\newblock \bibinfo{title}{{MIMO} transmission over a time-varying channel using
  {SVD}}.
\newblock \bibinfo{journal}{IEEE Trans. Wireless Commun.} \bibinfo{volume}{4},
  \bibinfo{pages}{757--764}.
\bibitem[{Lee et~al.(2017)Lee, Kim, Kim, Yoon and Choi}]{2017Deep}
\bibinfo{author}{Lee, J.H.}, \bibinfo{author}{Kim, B.}, \bibinfo{author}{Kim,
  J.}, \bibinfo{author}{Yoon, D.}, \bibinfo{author}{Choi, J.W.},
  \bibinfo{year}{2017}.
\newblock \bibinfo{title}{Deep neural network-based blind modulation
  classification for fading channels}, in: \bibinfo{booktitle}{Proc. Int. Conf.
  Inf. Commun. Technol. Converg.}
\bibitem[{Lee et~al.(2019)Lee, Kim and Shin}]{FPimages}
\bibinfo{author}{Lee, J.H.}, \bibinfo{author}{Kim, K.Y.},
  \bibinfo{author}{Shin, Y.}, \bibinfo{year}{2019}.
\newblock \bibinfo{title}{Feature image-based automatic modulation
  classification method using {CNN} algorithm}, in: \bibinfo{booktitle}{Proc.
  Int. Conf. Artif. Intell. Inf. Commun.}
\bibitem[{Li et~al.(2021)Li, Huang, Cheng, Meng and Han}]{9245503}
\bibinfo{author}{Li, L.}, \bibinfo{author}{Huang, J.}, \bibinfo{author}{Cheng,
  Q.}, \bibinfo{author}{Meng, H.}, \bibinfo{author}{Han, Z.},
  \bibinfo{year}{2021}.
\newblock \bibinfo{title}{Automatic modulation recognition: A few-shot learning
  method based on the capsule network}.
\newblock \bibinfo{journal}{IEEE Wireless Commun. Lett.} \bibinfo{volume}{10},
  \bibinfo{pages}{474--477}.
\bibitem[{Li et~al.(2019)Li, Shao and Wang}]{Bispectrum-Alexnet}
\bibinfo{author}{Li, Y.}, \bibinfo{author}{Shao, G.}, \bibinfo{author}{Wang,
  B.}, \bibinfo{year}{2019}.
\newblock \bibinfo{title}{Automatic modulation classification based on
  bispectrum and {CNN}}, in: \bibinfo{booktitle}{Proc. IEEE 8th Joint Int. Inf.
  Technol. Artif. Intell. Conf.}
\bibitem[{Liang et~al.(2021a)Liang, Tao, Wang, Su and Yang}]{9467342}
\bibinfo{author}{Liang, Z.}, \bibinfo{author}{Tao, M.}, \bibinfo{author}{Wang,
  L.}, \bibinfo{author}{Su, J.}, \bibinfo{author}{Yang, X.},
  \bibinfo{year}{2021}a.
\newblock \bibinfo{title}{Automatic modulation recognition based on adaptive
  attention mechanism and {ResNeXt} {WSL} model}.
\newblock \bibinfo{journal}{IEEE Commun. Lett.} \bibinfo{volume}{25},
  \bibinfo{pages}{2953--2957}.
\bibitem[{Liang et~al.(2021b)Liang, Wang, Tao, Xie and Yang}]{9682126}
\bibinfo{author}{Liang, Z.}, \bibinfo{author}{Wang, L.}, \bibinfo{author}{Tao,
  M.}, \bibinfo{author}{Xie, J.}, \bibinfo{author}{Yang, X.},
  \bibinfo{year}{2021}b.
\newblock \bibinfo{title}{Attention mechanism based {ResNeXt} network for
  automatic modulation classification}, in: \bibinfo{booktitle}{2021 IEEE
  Globecom Workshops}, pp. \bibinfo{pages}{1--6}.
\bibitem[{Lin et~al.(2022)Lin, Zeng and Gong}]{9672167}
\bibinfo{author}{Lin, S.}, \bibinfo{author}{Zeng, Y.}, \bibinfo{author}{Gong,
  Y.}, \bibinfo{year}{2022}.
\newblock \bibinfo{title}{Learning of time-frequency attention mechanism for
  automatic modulation recognition}.
\newblock \bibinfo{journal}{IEEE Wireless Commun. Lett.} ,
  \bibinfo{pages}{1--1}.
\bibitem[{Lin et~al.(2020)Lin, Tu and Dou}]{lin2020improved}
\bibinfo{author}{Lin, Y.}, \bibinfo{author}{Tu, Y.}, \bibinfo{author}{Dou, Z.},
  \bibinfo{year}{2020}.
\newblock \bibinfo{title}{An improved neural network pruning technology for
  automatic modulation classification in edge devices}.
\newblock \bibinfo{journal}{IEEE Trans. Veh. Technol.} \bibinfo{volume}{69},
  \bibinfo{pages}{5703--5706}.
\bibitem[{Liu and Xu(2006)}]{liu2006novel}
\bibinfo{author}{Liu, L.}, \bibinfo{author}{Xu, J.}, \bibinfo{year}{2006}.
\newblock \bibinfo{title}{A novel modulation classification method based on
  high order cumulants}, in: \bibinfo{booktitle}{Proc. Int. Conf. Wireless
  Commun., Netw. Mobile Comput.}, pp. \bibinfo{pages}{1--5}.
\bibitem[{Liu et~al.(2017)Liu, Yang and El~Gamal}]{liu2017deep}
\bibinfo{author}{Liu, X.}, \bibinfo{author}{Yang, D.},
  \bibinfo{author}{El~Gamal, A.}, \bibinfo{year}{2017}.
\newblock \bibinfo{title}{Deep neural network architectures for modulation
  classification}, in: \bibinfo{booktitle}{Proc. 51st Asilomar Conf. Signals,
  Syst., Comput.}, pp. \bibinfo{pages}{915--919}.
\bibitem[{Ma et~al.(2020)Ma, Xu, Meng, Wang and Wang}]{2020Cross}
\bibinfo{author}{Ma, H.}, \bibinfo{author}{Xu, G.}, \bibinfo{author}{Meng, H.},
  \bibinfo{author}{Wang, M.}, \bibinfo{author}{Wang, W.}, \bibinfo{year}{2020}.
\newblock \bibinfo{title}{Cross model deep learning scheme for automatic
  modulation classification}.
\newblock \bibinfo{journal}{IEEE Access} \bibinfo{volume}{PP},
  \bibinfo{pages}{1--1}.
\bibitem[{Ma et~al.(2019)Ma, Lin, Gao and Qiu}]{CCESRESTNET}
\bibinfo{author}{Ma, J.}, \bibinfo{author}{Lin, S.C.}, \bibinfo{author}{Gao,
  H.}, \bibinfo{author}{Qiu, T.}, \bibinfo{year}{2019}.
\newblock \bibinfo{title}{Automatic modulation classification under
  non-gaussian noise: A deep residual learning approach}, in:
  \bibinfo{booktitle}{Proc. IEEE Int. Conf. Commun.}
\bibitem[{Mendis et~al.(2016)Mendis, Wei and Madanayake}]{mendis2016deep}
\bibinfo{author}{Mendis, G.J.}, \bibinfo{author}{Wei, J.},
  \bibinfo{author}{Madanayake, A.}, \bibinfo{year}{2016}.
\newblock \bibinfo{title}{Deep learning-based automated modulation
  classification for cognitive radio}, in: \bibinfo{booktitle}{Proc. IEEE Int.
  Conf. Commun. Syst.}, pp. \bibinfo{pages}{1--6}.
\bibitem[{Mendis et~al.(2019)Mendis, Wei-Kocsis and
  Madanayake}]{mendis2019deep}
\bibinfo{author}{Mendis, G.J.}, \bibinfo{author}{Wei-Kocsis, J.},
  \bibinfo{author}{Madanayake, A.}, \bibinfo{year}{2019}.
\newblock \bibinfo{title}{Deep learning based radio-signal identification with
  hardware design}.
\newblock \bibinfo{journal}{IEEE Trans. Aerosp. Electron. Syst.}
  \bibinfo{volume}{55}, \bibinfo{pages}{2516--2531}.
\bibitem[{Njoku et~al.(2021)Njoku, Morocho-Cayamcela and Lim}]{9349627}
\bibinfo{author}{Njoku, J.N.}, \bibinfo{author}{Morocho-Cayamcela, M.E.},
  \bibinfo{author}{Lim, W.}, \bibinfo{year}{2021}.
\newblock \bibinfo{title}{{CGDNet}: Efficient hybrid deep learning model for
  robust automatic modulation recognition}.
\newblock \bibinfo{journal}{IEEE Netw. Lett.} \bibinfo{volume}{3},
  \bibinfo{pages}{47--51}.
\bibitem[{O'shea and West(2016)}]{o2016radio}
\bibinfo{author}{O'shea, T.J.}, \bibinfo{author}{West, N.},
  \bibinfo{year}{2016}.
\newblock \bibinfo{title}{Radio machine learning dataset generation with gnu
  radio}, in: \bibinfo{booktitle}{Proc. GNU Radio Conf.}
\bibitem[{O’Shea et~al.(2016)O’Shea, Corgan and
  Clancy}]{o2016convolutional}
\bibinfo{author}{O’Shea, T.J.}, \bibinfo{author}{Corgan, J.},
  \bibinfo{author}{Clancy, T.C.}, \bibinfo{year}{2016}.
\newblock \bibinfo{title}{Convolutional radio modulation recognition networks},
  in: \bibinfo{booktitle}{Proc. Int. Conf. Eng. Appl. Neural Netw.},
  \bibinfo{organization}{Springer}. pp. \bibinfo{pages}{213--226}.
\bibitem[{O’Shea et~al.(2018)O’Shea, Roy and Clancy}]{o2018over}
\bibinfo{author}{O’Shea, T.J.}, \bibinfo{author}{Roy, T.},
  \bibinfo{author}{Clancy, T.C.}, \bibinfo{year}{2018}.
\newblock \bibinfo{title}{Over-the-air deep learning based radio signal
  classification}.
\newblock \bibinfo{journal}{IEEE J. Sel. Top. Signal Process.}
  \bibinfo{volume}{12}, \bibinfo{pages}{168--179}.
\bibitem[{Park et~al.(2008)Park, Choi, Nah, Jang and Kim}]{park2008automatic}
\bibinfo{author}{Park, C.S.}, \bibinfo{author}{Choi, J.H.},
  \bibinfo{author}{Nah, S.P.}, \bibinfo{author}{Jang, W.},
  \bibinfo{author}{Kim, D.Y.}, \bibinfo{year}{2008}.
\newblock \bibinfo{title}{Automatic modulation recognition of digital signals
  using wavelet features and {SVM}}, in: \bibinfo{booktitle}{Proc. Int. Conf.
  Adv. Commun. Technol.}, pp. \bibinfo{pages}{387--390}.
\bibitem[{Peng et~al.(2018)Peng, Jiang, Wang, Alwageed and
  Yao}]{2018Modulation}
\bibinfo{author}{Peng, S.}, \bibinfo{author}{Jiang, H.}, \bibinfo{author}{Wang,
  H.}, \bibinfo{author}{Alwageed, H.}, \bibinfo{author}{Yao, Y.D.},
  \bibinfo{year}{2018}.
\newblock \bibinfo{title}{Modulation classification based on signal
  constellation diagrams and deep learning}.
\newblock \bibinfo{journal}{IEEE Trans. Neural Netw. Learn. Syst.}
  \bibinfo{volume}{PP}, \bibinfo{pages}{1--10}.
\bibitem[{Qi et~al.(2020)Qi, Zhou, Zheng and Li}]{AMCmulti}
\bibinfo{author}{Qi, P.}, \bibinfo{author}{Zhou, X.}, \bibinfo{author}{Zheng,
  S.}, \bibinfo{author}{Li, Z.}, \bibinfo{year}{2020}.
\newblock \bibinfo{title}{Automatic modulation classification based on deep
  residual networks with multimodal information}.
\newblock \bibinfo{journal}{IEEE Trans. Cogn. Commun. Netw.}
  \bibinfo{volume}{PP}, \bibinfo{pages}{1--1}.
\bibitem[{Rajendran et~al.(2018)Rajendran, Meert, Giustiniano, Lenders and
  Pollin}]{rajendran2018deep}
\bibinfo{author}{Rajendran, S.}, \bibinfo{author}{Meert, W.},
  \bibinfo{author}{Giustiniano, D.}, \bibinfo{author}{Lenders, V.},
  \bibinfo{author}{Pollin, S.}, \bibinfo{year}{2018}.
\newblock \bibinfo{title}{Deep learning models for wireless signal
  classification with distributed low-cost spectrum sensors}.
\newblock \bibinfo{journal}{IEEE Trans. Cogn. Commun. Netw.}
  \bibinfo{volume}{4}, \bibinfo{pages}{433--445}.
\bibitem[{Ramjee et~al.(2021)Ramjee, Ju, Yang, Liu, Gamal and Eldar}]{9525144}
\bibinfo{author}{Ramjee, S.}, \bibinfo{author}{Ju, S.}, \bibinfo{author}{Yang,
  D.}, \bibinfo{author}{Liu, X.}, \bibinfo{author}{Gamal, A.E.},
  \bibinfo{author}{Eldar, Y.C.}, \bibinfo{year}{2021}.
\newblock \bibinfo{title}{Ensemble wrapper subsampling for deep modulation
  classification}.
\newblock \bibinfo{journal}{IEEE Trans. Cogn. Commun. Netw.}
  \bibinfo{volume}{7}, \bibinfo{pages}{1156--1170}.
\bibitem[{Sahay et~al.(2021)Sahay, Brinton and Love}]{9542973}
\bibinfo{author}{Sahay, R.}, \bibinfo{author}{Brinton, C.G.},
  \bibinfo{author}{Love, D.J.}, \bibinfo{year}{2021}.
\newblock \bibinfo{title}{A deep ensemble-based wireless receiver architecture
  for mitigating adversarial attacks in automatic modulation classification}.
\newblock \bibinfo{journal}{IEEE Trans. Cogn. Commun. Netw.} ,
  \bibinfo{pages}{1--1}.
\bibitem[{Shah and Dang(2020)}]{STBCMIMO}
\bibinfo{author}{Shah, M.H.}, \bibinfo{author}{Dang, X.}, \bibinfo{year}{2020}.
\newblock \bibinfo{title}{Low-complexity deep learning and {RBFN} architectures
  for modulation classification of space-time block-code {(STBC)-MIMO}
  systems}.
\newblock \bibinfo{journal}{Digit. Signal Process.} \bibinfo{volume}{99},
  \bibinfo{pages}{102656}.
\bibitem[{Shi et~al.(2022)Shi, Hu, Yue and Shen}]{shi2022combining}
\bibinfo{author}{Shi, F.}, \bibinfo{author}{Hu, Z.}, \bibinfo{author}{Yue, C.},
  \bibinfo{author}{Shen, Z.}, \bibinfo{year}{2022}.
\newblock \bibinfo{title}{Combining neural networks for modulation
  recognition}.
\newblock \bibinfo{journal}{Digit. Signal Process.} \bibinfo{volume}{120},
  \bibinfo{pages}{103264}.
\bibitem[{Shi et~al.(2019)Shi, Liu, Cheng, Li and Zhao}]{2019Particle}
\bibinfo{author}{Shi, W.}, \bibinfo{author}{Liu, D.}, \bibinfo{author}{Cheng,
  X.}, \bibinfo{author}{Li, Y.}, \bibinfo{author}{Zhao, Y.},
  \bibinfo{year}{2019}.
\newblock \bibinfo{title}{Particle swarm optimization-based deep neural network
  for digital modulation recognition}.
\newblock \bibinfo{journal}{IEEE Access} \bibinfo{volume}{7},
  \bibinfo{pages}{104591--104600}.
\bibitem[{Swami and Sadler(2000)}]{swami2000hierarchical}
\bibinfo{author}{Swami, A.}, \bibinfo{author}{Sadler, B.M.},
  \bibinfo{year}{2000}.
\newblock \bibinfo{title}{Hierarchical digital modulation classification using
  cumulants}.
\newblock \bibinfo{journal}{IEEE Trans. commun.} \bibinfo{volume}{48},
  \bibinfo{pages}{416--429}.
\bibitem[{Tang et~al.(2018)Tang, Tu, Zhang and Yun}]{2018Digital}
\bibinfo{author}{Tang, B.}, \bibinfo{author}{Tu, Y.}, \bibinfo{author}{Zhang,
  S.}, \bibinfo{author}{Yun, L.}, \bibinfo{year}{2018}.
\newblock \bibinfo{title}{Digital signal modulation classification with data
  augmentation using generative adversarial nets in cognitive radio networks}.
\newblock \bibinfo{journal}{IEEE Access} \bibinfo{volume}{PP},
  \bibinfo{pages}{1--1}.
\bibitem[{Tekb{\i}y{\i}k et~al.(2020)Tekb{\i}y{\i}k, Ekti, G{\"o}r{\c{c}}in,
  Kurt and Ke{\c{c}}eci}]{tekbiyik2020robust}
\bibinfo{author}{Tekb{\i}y{\i}k, K.}, \bibinfo{author}{Ekti, A.R.},
  \bibinfo{author}{G{\"o}r{\c{c}}in, A.}, \bibinfo{author}{Kurt, G.K.},
  \bibinfo{author}{Ke{\c{c}}eci, C.}, \bibinfo{year}{2020}.
\newblock \bibinfo{title}{Robust and fast automatic modulation classification
  with {CNN} under multipath fading channels}, in: \bibinfo{booktitle}{Proc.
  IEEE 91st Veh. Technol. Conf.}, pp. \bibinfo{pages}{1--6}.
\bibitem[{Tu and Lin(2019)}]{accesspruning}
\bibinfo{author}{Tu, Y.}, \bibinfo{author}{Lin, Y.}, \bibinfo{year}{2019}.
\newblock \bibinfo{title}{Deep neural network compression technique towards
  efficient digital signal modulation recognition in edge device}.
\newblock \bibinfo{journal}{IEEE Access} \bibinfo{volume}{7},
  \bibinfo{pages}{58113--58119}.
\bibitem[{Tugnait(2001)}]{tugnait2001blind}
\bibinfo{author}{Tugnait, J.K.}, \bibinfo{year}{2001}.
\newblock \bibinfo{title}{Blind estimation and equalization of {MIMO} channels
  via multidelay whitening}.
\newblock \bibinfo{journal}{IEEE J. Sel. Areas Commun.} \bibinfo{volume}{19},
  \bibinfo{pages}{1507--1519}.
\bibitem[{Wang et~al.(2017)Wang, Zhang, Li, Li, Fu, Cui and Chen}]{eyediagram}
\bibinfo{author}{Wang, D.}, \bibinfo{author}{Zhang, M.}, \bibinfo{author}{Li,
  Z.}, \bibinfo{author}{Li, J.}, \bibinfo{author}{Fu, M.},
  \bibinfo{author}{Cui, Y.}, \bibinfo{author}{Chen, X.}, \bibinfo{year}{2017}.
\newblock \bibinfo{title}{Modulation format recognition and {OSNR} estimation
  using {CNN}-based deep learning}.
\newblock \bibinfo{journal}{IEEE Photon. Technol. Lett.} \bibinfo{volume}{29},
  \bibinfo{pages}{1667--1670}.
\bibitem[{Wang et~al.(2021)Wang, Hou, Zhang and Guo}]{9427151}
\bibinfo{author}{Wang, T.}, \bibinfo{author}{Hou, Y.}, \bibinfo{author}{Zhang,
  H.}, \bibinfo{author}{Guo, Z.}, \bibinfo{year}{2021}.
\newblock \bibinfo{title}{Deep learning based modulation recognition with
  multi-cue fusion}.
\newblock \bibinfo{journal}{IEEE Wireless Commun. Lett.} \bibinfo{volume}{10},
  \bibinfo{pages}{1757--1760}.
\bibitem[{Wang et~al.(2020a)Wang, Gui, Gacanin, Ohtsuki, Sari and
  Adachi}]{wang2020transfer}
\bibinfo{author}{Wang, Y.}, \bibinfo{author}{Gui, G.},
  \bibinfo{author}{Gacanin, H.}, \bibinfo{author}{Ohtsuki, T.},
  \bibinfo{author}{Sari, H.}, \bibinfo{author}{Adachi, F.},
  \bibinfo{year}{2020}a.
\newblock \bibinfo{title}{Transfer learning for semi-supervised automatic
  modulation classification in {ZF-MIMO} systems}.
\newblock \bibinfo{journal}{IEEE Trans. Emerg. Sel. Topics Circuits Syst.}
  \bibinfo{volume}{10}, \bibinfo{pages}{231--239}.
\bibitem[{Wang et~al.(2020b)Wang, Gui, Yin, Wang, Sun, Gui, Gacanin, Sari and
  Adachi}]{wang2020automatic}
\bibinfo{author}{Wang, Y.}, \bibinfo{author}{Gui, J.}, \bibinfo{author}{Yin,
  Y.}, \bibinfo{author}{Wang, J.}, \bibinfo{author}{Sun, J.},
  \bibinfo{author}{Gui, G.}, \bibinfo{author}{Gacanin, H.},
  \bibinfo{author}{Sari, H.}, \bibinfo{author}{Adachi, F.},
  \bibinfo{year}{2020}b.
\newblock \bibinfo{title}{Automatic modulation classification for {MIMO}
  systems via deep learning and zero-forcing equalization}.
\newblock \bibinfo{journal}{IEEE Trans. Veh. Technol.} \bibinfo{volume}{69},
  \bibinfo{pages}{5688--5692}.
\bibitem[{Wang et~al.(2019)Wang, Liu, Yang and Gui}]{wang2019data}
\bibinfo{author}{Wang, Y.}, \bibinfo{author}{Liu, M.}, \bibinfo{author}{Yang,
  J.}, \bibinfo{author}{Gui, G.}, \bibinfo{year}{2019}.
\newblock \bibinfo{title}{Data-driven deep learning for automatic modulation
  recognition in cognitive radios}.
\newblock \bibinfo{journal}{IEEE Trans. Veh. Technol.} \bibinfo{volume}{68},
  \bibinfo{pages}{4074--4077}.
\bibitem[{Wang et~al.(2020c)Wang, Wang, Zhang, Yang and Gui}]{wang2020deep}
\bibinfo{author}{Wang, Y.}, \bibinfo{author}{Wang, J.}, \bibinfo{author}{Zhang,
  W.}, \bibinfo{author}{Yang, J.}, \bibinfo{author}{Gui, G.},
  \bibinfo{year}{2020}c.
\newblock \bibinfo{title}{Deep learning-based cooperative automatic modulation
  classification method for {MIMO} systems}.
\newblock \bibinfo{journal}{IEEE Trans. Veh. Technol.} \bibinfo{volume}{69},
  \bibinfo{pages}{4575--4579}.
\bibitem[{Wang et~al.(2020d)Wang, Yang, Liu and Gui}]{wang2020lightamc}
\bibinfo{author}{Wang, Y.}, \bibinfo{author}{Yang, J.}, \bibinfo{author}{Liu,
  M.}, \bibinfo{author}{Gui, G.}, \bibinfo{year}{2020}d.
\newblock \bibinfo{title}{Lightamc: Lightweight automatic modulation
  classification via deep learning and compressive sensing}.
\newblock \bibinfo{journal}{IEEE Trans. Veh. Technol.} \bibinfo{volume}{69},
  \bibinfo{pages}{3491--3495}.
\bibitem[{Wei and Mendel(2000)}]{wei2000maximum}
\bibinfo{author}{Wei, W.}, \bibinfo{author}{Mendel, J.M.},
  \bibinfo{year}{2000}.
\newblock \bibinfo{title}{Maximum-likelihood classification for digital
  amplitude-phase modulations}.
\newblock \bibinfo{journal}{IEEE Trans. Commun.} \bibinfo{volume}{48},
  \bibinfo{pages}{189--193}.
\bibitem[{West and O'Shea(2017)}]{west2017deep}
\bibinfo{author}{West, N.E.}, \bibinfo{author}{O'Shea, T.},
  \bibinfo{year}{2017}.
\newblock \bibinfo{title}{Deep architectures for modulation recognition}, in:
  \bibinfo{booktitle}{Proc. IEEE Int. Symp. Dyn. Spectr. Access Netw.,}, pp.
  \bibinfo{pages}{1--6}.
\bibitem[{Wong and Nandi(2004)}]{2004Automatic}
\bibinfo{author}{Wong, M.}, \bibinfo{author}{Nandi, A.K.},
  \bibinfo{year}{2004}.
\newblock \bibinfo{title}{Automatic digital modulation recognition using
  artificial neural network and genetic algorithm}.
\newblock \bibinfo{journal}{Signal Process.} \bibinfo{volume}{84},
  \bibinfo{pages}{351--365}.
\bibitem[{Wu et~al.(2019)Wu, Li, Zhou and Meng}]{Multifeafusion}
\bibinfo{author}{Wu, H.}, \bibinfo{author}{Li, Y.}, \bibinfo{author}{Zhou, L.},
  \bibinfo{author}{Meng, J.}, \bibinfo{year}{2019}.
\newblock \bibinfo{title}{Convolutional neural network and multi-feature fusion
  for automatic modulation classification}.
\newblock \bibinfo{journal}{Electron. Lett.} \bibinfo{volume}{55},
  \bibinfo{pages}{895--897}.
\bibitem[{Xie et~al.(2019)Xie, Hu, Yu, Zhu and Ouyang}]{2019Deepdnn}
\bibinfo{author}{Xie, W.}, \bibinfo{author}{Hu, S.}, \bibinfo{author}{Yu, C.},
  \bibinfo{author}{Zhu, P.}, \bibinfo{author}{Ouyang, J.},
  \bibinfo{year}{2019}.
\newblock \bibinfo{title}{Deep learning in digital modulation recognition using
  high order cumulants}.
\newblock \bibinfo{journal}{IEEE Access} \bibinfo{volume}{PP},
  \bibinfo{pages}{1--1}.
\bibitem[{Xu et~al.(2020)Xu, Luo, Parr and Luo}]{xu2020spatiotemporal}
\bibinfo{author}{Xu, J.}, \bibinfo{author}{Luo, C.}, \bibinfo{author}{Parr,
  G.}, \bibinfo{author}{Luo, Y.}, \bibinfo{year}{2020}.
\newblock \bibinfo{title}{A spatiotemporal multi-channel learning framework for
  automatic modulation recognition}.
\newblock \bibinfo{journal}{IEEE Wireless Commun. Lett.} .
\bibitem[{Xu et~al.(2010)Xu, Su and Zhou}]{xu2010likelihood}
\bibinfo{author}{Xu, J.L.}, \bibinfo{author}{Su, W.}, \bibinfo{author}{Zhou,
  M.}, \bibinfo{year}{2010}.
\newblock \bibinfo{title}{Likelihood-ratio approaches to automatic modulation
  classification}.
\newblock \bibinfo{journal}{IEEE Trans. Syst. Man Cybern. C}
  \bibinfo{volume}{41}, \bibinfo{pages}{455--469}.
\bibitem[{Xu et~al.(2008)Xu, Ge and Wang}]{xu2008digital}
\bibinfo{author}{Xu, Y.}, \bibinfo{author}{Ge, L.}, \bibinfo{author}{Wang, B.},
  \bibinfo{year}{2008}.
\newblock \bibinfo{title}{Digital modulation recognition method based on
  self-organizing map neural networks}, in: \bibinfo{booktitle}{Proc. Int.
  Conf. Wireless Commun., Netw. Mobile Comput.}, pp. \bibinfo{pages}{1--4}.
\bibitem[{Yashashwi et~al.(2018)Yashashwi, Sethi and
  Chaporkar}]{yashashwi2018learnable}
\bibinfo{author}{Yashashwi, K.}, \bibinfo{author}{Sethi, A.},
  \bibinfo{author}{Chaporkar, P.}, \bibinfo{year}{2018}.
\newblock \bibinfo{title}{A learnable distortion correction module for
  modulation recognition}.
\newblock \bibinfo{journal}{IEEE Wireless Commun. Lett.} \bibinfo{volume}{8},
  \bibinfo{pages}{77--80}.
\bibitem[{Yuan et~al.(2004)Yuan, Zhao-Yang and Pei-Liang}]{yuan2004modulation}
\bibinfo{author}{Yuan, J.}, \bibinfo{author}{Zhao-Yang, Z.},
  \bibinfo{author}{Pei-Liang, Q.}, \bibinfo{year}{2004}.
\newblock \bibinfo{title}{Modulation classification of communication signals},
  in: \bibinfo{booktitle}{Proc. IEEE Military Commun. Conf.}, pp.
  \bibinfo{pages}{1470--1476}.
\bibitem[{Zeng and Ng(2006)}]{zeng2006blind}
\bibinfo{author}{Zeng, Y.}, \bibinfo{author}{Ng, T.S.}, \bibinfo{year}{2006}.
\newblock \bibinfo{title}{Blind estimation of {MIMO} channels with an upper
  bound for channel orders}.
\newblock \bibinfo{journal}{Signal Process.} \bibinfo{volume}{86},
  \bibinfo{pages}{212--222}.
\bibitem[{Zeng et~al.(2019)Zeng, Zhang, Han, Gong and Zhang}]{zeng2019spectrum}
\bibinfo{author}{Zeng, Y.}, \bibinfo{author}{Zhang, M.}, \bibinfo{author}{Han,
  F.}, \bibinfo{author}{Gong, Y.}, \bibinfo{author}{Zhang, J.},
  \bibinfo{year}{2019}.
\newblock \bibinfo{title}{Spectrum analysis and convolutional neural network
  for automatic modulation recognition}.
\newblock \bibinfo{journal}{IEEE Wireless Commun. Lett.} \bibinfo{volume}{8},
  \bibinfo{pages}{929--932}.
\bibitem[{Zhang et~al.(2021a)Zhang, Luo, Xu and Luo}]{zhang2021efficient}
\bibinfo{author}{Zhang, F.}, \bibinfo{author}{Luo, C.}, \bibinfo{author}{Xu,
  J.}, \bibinfo{author}{Luo, Y.}, \bibinfo{year}{2021}a.
\newblock \bibinfo{title}{An efficient deep learning model for automatic
  modulation recognition based on parameter estimation and transformation}.
\newblock \bibinfo{journal}{IEEE Commun. Lett.} \bibinfo{volume}{25},
  \bibinfo{pages}{3287--3290}.
\bibitem[{Zhang et~al.(2021b)Zhang, Lin, Yan, Ling and Wang}]{9467343}
\bibinfo{author}{Zhang, L.}, \bibinfo{author}{Lin, C.}, \bibinfo{author}{Yan,
  W.}, \bibinfo{author}{Ling, Q.}, \bibinfo{author}{Wang, Y.},
  \bibinfo{year}{2021}b.
\newblock \bibinfo{title}{Real-time {OFDM} signal modulation classification
  based on deep learning and software-defined radio}.
\newblock \bibinfo{journal}{IEEE Commun. Let.} \bibinfo{volume}{25},
  \bibinfo{pages}{2988--2992}.
\bibitem[{Zhang et~al.(2021c)Zhang, Liu, Yang, Jiang and Wu}]{9440746}
\bibinfo{author}{Zhang, L.}, \bibinfo{author}{Liu, H.}, \bibinfo{author}{Yang,
  X.}, \bibinfo{author}{Jiang, Y.}, \bibinfo{author}{Wu, Z.},
  \bibinfo{year}{2021}c.
\newblock \bibinfo{title}{Intelligent denoising-aided deep learning modulation
  recognition with cyclic spectrum features for higher accuracy}.
\newblock \bibinfo{journal}{IEEE Trans. Aerosp. Electron. Syst.}
  \bibinfo{volume}{57}, \bibinfo{pages}{3749--3757}.
\bibitem[{Zhang et~al.(2018)Zhang, Zeng, Han and Gong}]{IQFOC}
\bibinfo{author}{Zhang, M.}, \bibinfo{author}{Zeng, Y.}, \bibinfo{author}{Han,
  Z.}, \bibinfo{author}{Gong, Y.}, \bibinfo{year}{2018}.
\newblock \bibinfo{title}{Automatic modulation recognition using deep learning
  architectures}, in: \bibinfo{booktitle}{Proc. IEEE 19th Int. Workshop Signal
  Process. Adv. Wireless Commun.}, pp. \bibinfo{pages}{1--5}.
\bibitem[{Zhang et~al.(2021d)Zhang, Feng, Krunz and Hossein
  Yazdani~Abyaneh}]{9488834}
\bibinfo{author}{Zhang, W.}, \bibinfo{author}{Feng, M.},
  \bibinfo{author}{Krunz, M.}, \bibinfo{author}{Hossein Yazdani~Abyaneh, A.},
  \bibinfo{year}{2021}d.
\newblock \bibinfo{title}{Signal detection and classification in shared
  spectrum: A deep learning approach}, in: \bibinfo{booktitle}{IEEE Conf.
  Comput. Commun.}, pp. \bibinfo{pages}{1--10}.
\bibitem[{Zhang et~al.(2022)Zhang, Li, Zhai, Li and Gao}]{9676419}
\bibinfo{author}{Zhang, Z.}, \bibinfo{author}{Li, Y.}, \bibinfo{author}{Zhai,
  Q.}, \bibinfo{author}{Li, Y.}, \bibinfo{author}{Gao, M.},
  \bibinfo{year}{2022}.
\newblock \bibinfo{title}{Few-shot learning for fine-grained signal modulation
  recognition based on foreground segmentation}.
\newblock \bibinfo{journal}{IEEE Trans. Veh. Technol.} , \bibinfo{pages}{1--1}.
\bibitem[{Zhang et~al.(2020)Zhang, Luo, Wang, Gan and
  Xiang}]{zhang2020automatic}
\bibinfo{author}{Zhang, Z.}, \bibinfo{author}{Luo, H.}, \bibinfo{author}{Wang,
  C.}, \bibinfo{author}{Gan, C.}, \bibinfo{author}{Xiang, Y.},
  \bibinfo{year}{2020}.
\newblock \bibinfo{title}{Automatic modulation classification using {CNN-LSTM}
  based dual-stream structure}.
\newblock \bibinfo{journal}{IEEE Trans. Veh. Technol.} \bibinfo{volume}{69},
  \bibinfo{pages}{13521--13531}.
\bibitem[{Zhang et~al.(2019)Zhang, Wang, Gan, Sun and Wang}]{BJD}
\bibinfo{author}{Zhang, Z.}, \bibinfo{author}{Wang, C.}, \bibinfo{author}{Gan,
  C.}, \bibinfo{author}{Sun, S.}, \bibinfo{author}{Wang, M.},
  \bibinfo{year}{2019}.
\newblock \bibinfo{title}{Automatic modulation classification using
  convolutional neural network with features fusion of {SPWVD} and {BJD}}.
\newblock \bibinfo{journal}{IEEE Trans. Signal Inf. Process. Netw.} ,
  \bibinfo{pages}{1--1}.
\bibitem[{Zhou et~al.(2021)Zhou, Zhang, Mu, Zhang, Zhang and Jing}]{9650842}
\bibinfo{author}{Zhou, Q.}, \bibinfo{author}{Zhang, R.}, \bibinfo{author}{Mu,
  J.}, \bibinfo{author}{Zhang, H.}, \bibinfo{author}{Zhang, F.},
  \bibinfo{author}{Jing, X.}, \bibinfo{year}{2021}.
\newblock \bibinfo{title}{{AMCRN}: Few-shot learning for automatic modulation
  classification}.
\newblock \bibinfo{journal}{IEEE Commun. Lett.} , \bibinfo{pages}{1--1}.

\end{thebibliography}






\end{document}